\documentclass[numberedappendix]{emulateapj}

\usepackage{subfigure}

\begin{document}
\slugcomment{{\sc Accepted to ApJ:} October 29, 2008}

\def\s{{\rm\,s}} 
\def\sr{{\rm\,sr}} 
\def\erg{{\rm\,erg}} 
\def\cm{{\rm\,cm}} 
\def\m{{\rm\,m}} 
\def\km{{\rm\,km}} 
\def\mm{{\rm\,mm}} 
\def\gm{{\rm\,g}} 
\def\g{{\rm\,g}} 
\def\kg{{\rm\,kg}} 
\def\au{{\rm AU}} 
\def\deg{{\rm deg}} 
\def\rad{{\rm rad}}   
\def\AU{{\rm\, AU}}  
\def\K{{\rm\,K}}  
\def\yr{{\rm\,yr}}  
\def\Hz{{\rm\,Hz}}  
\def\eV{{\rm\,eV}}  
\def\baru{{\rm\,bar}}

\def\ion#1#2{#1$\;${\small\rm\@{#2}}\relax}

\title{Atmospheric Escape from Hot Jupiters}
\author{Ruth A.~Murray-Clay\altaffilmark{1,2}, Eugene I.~Chiang\altaffilmark{1}, \& Norman Murray\altaffilmark{3,4}}
\altaffiltext{1}{Center for Integrative Planetary Sciences,
Astronomy Department,
University of California at Berkeley,
Berkeley, CA~94720, USA}
\altaffiltext{2}{Harvard-Smithsonian Center for Astrophysics, 60 Garden Street, MS-51, Cambridge, MA~02138, USA}
\altaffiltext{3}{Canadian Institute for Theoretical Astrophysics, University of Toronto,
Toronto, ON M5S 3H8, Canada}
\altaffiltext{4}{Canada Research Chair in Astrophysics}

\email{rmurray-clay@cfa.harvard.edu}

\begin{abstract}

  Photoionization heating from UV radiation incident on the
  atmospheres of hot Jupiters may drive planetary mass
  loss. Observations of stellar Lyman-$\alpha$ absorption have
  suggested that the hot Jupiter HD 209458b is losing atomic hydrogen.
  We construct a model of escape that includes realistic heating and
  cooling, ionization balance, tidal gravity, and pressure confinement
  by the host star wind. We show that mass loss takes the form of a
  hydrodynamic (``Parker'') wind, emitted from the planet's dayside
  during lulls in the stellar wind.  When dayside winds are suppressed
  by the confining action of the stellar wind, nightside winds might
  pick up if there is sufficient horizontal transport of heat.
A hot Jupiter loses mass at maximum rates of $\sim$$2 \times
10^{12}\gm \s^{-1}$ during its host star's pre-main-sequence phase and
$\sim$$2 \times10^{10}\gm \s^{-1}$ during the star's main sequence
lifetime, for total maximum losses of $\sim$0.06\% and $\sim$0.6\% of
the planet's mass, respectively.  
For UV fluxes $F_{\rm UV} \lesssim
10^4 \erg \cm^{-2} \s^{-1}$, the mass loss rate is approximately
energy-limited and scales as $\dot{M} \propto F_{\rm UV}^{0.9}$. For
larger UV fluxes, such as those typical of T Tauri stars, radiative
losses and plasma recombination force $\dot{M}$ to increase more
slowly as $F_{\rm UV}^{0.6}$. Dayside winds are quenched during the T
Tauri phase because of confinement by overwhelming stellar wind
pressure. During this early stage, nightside winds can still blow if
the planet resides outside the stellar Alfv\'en radius; otherwise,
even nightside winds are stifled by stellar magnetic pressure, and
mass loss is restricted to polar regions.  
We conclude that while
UV radiation can indeed drive winds from hot Jupiters,
such winds cannot significantly alter planetary masses during any evolutionary
stage.  They can, however, produce observable signatures.  Candidates for explaining why the Lyman-$\alpha$ photons of HD 209458 are absorbed at Doppler-shifted velocities of $\pm$100 km/s include charge-exchange in the shock between the planetary and stellar winds.
\end{abstract}


\section{INTRODUCTION}\label{sec-intro}
About 1/5 of the approximately 200 extrasolar planets discovered to
date have masses comparable to Jupiter's, but orbit their host stars
at distances less than 0.1 AU \citep[e.g.,][]{bwm+06}.  Since stellar
heating inhibits the formation of gas giants around Sun-like stars at
such distances \citep[e.g.,][]{r06}, these ``hot Jupiters'' likely
migrated inward by disk torques from where they were born \citep[e.g.,][and references therein]{pnk+07}.

Once parked (possibly because the disk was truncated
by the stellar magnetosphere; 
\citealt{lbr96}), hot Jupiters are bathed in ultraviolet (UV) radiation
from their host stars.  Atmospheric gas is heated by photoionization
of hydrogen, and escapes.
Several groups have argued that the outflows can
evaporate gas giants nearly entirely,
laying bare their rocky cores \citep{lsr+03,
  bsc+04, bcb+05}.  In fact, hot Jupiters are observed to have
systematically lower masses than extrasolar planets at larger
distances from their stars \citep{zm02}.  However,
\citet{hhb+07,hhb+07a} are unable to reproduce the mass distribution
of hot Jupiters by experimentation with mass loss histories.

Observations of HD 209458b, the first hot Jupiter observed to transit
its host star \citep{hmb+00, cbl+00}, suggest that the planet may indeed be
losing atomic hydrogen.  \citet{vld+03} used the Hubble
Space Telescope Imaging Spectrograph (STIS) in a high spectral
resolution mode to measure Lyman-$\alpha$ emission from HD 209458b's
host star in and out of transit.  They observed, with $3\sigma$
confidence, the Ly$\alpha$ flux---at wavelengths shifted from line
center by Doppler equivalent velocities of $\pm 100 \km \s^{-1}$---to
decrease during transit by $\sim$15\% (as integrated over Doppler
equivalent velocities extending approximately from -50 to -140 km/s
and from +30 to +100 km/s).  The authors attributed this
decrease to absorption by intervening atomic hydrogen surrounding the
planet: a halo of gas sufficiently distended and traveling at fast
enough velocities to no longer be bound to the planet.  The absorption
signal was reported to be stronger at blueshifted wavelengths than at
redshifted wavelengths; blueshifted velocities (towards the observer,
away from the star) were argued to arise from stellar radiation
pressure acting on gas via the Ly$\alpha$ line.

\citet{b07} reanalyzed the STIS data and agreed that stellar
Ly$\alpha$ photons were absorbed by planetary gas during transit, but
found the signal to be weaker: the flux decreased by only
$8.9 \pm 2.1$\%, at similar Doppler equivalent velocities of $\pm$100
km/s. Moreover, no preference for blueshifted absorption was
found. \citet{b07} cautioned that intrinsic stellar variability could
easily produce a spurious preference for either negative or positive
velocities.  Despite this and other complications, \citet{b07} decided
nonetheless that the absorption signal could be of planetary
origin. From the effective occulting area corresponding to an 8.9\%
drop in flux, it was concluded that the obscuring hydrogen, while
occupying a ``corona'' significantly more inflated than the visible
photosphere, remained bound within the planet's Roche lobe. This last
argument, as pointed out by \citet{vld+08}, is not secure.  If the
wavelength shifts associated with the putative absorption reflect
Doppler shifts from bulk flows, then the relevant velocities,
regardless of whether they are positive or negative, are larger than
the planet's escape velocity and cannot arise from a bound hydrogen
atmosphere.  Furthermore, as a point of principle, gas can flow past
the Roche lobe of the planet and elude detection if it is optically
thin to Ly$\alpha$ photons; therefore arguments based on
effective occulting area are not conclusive.

Certainly we share the concern of \citet{b07} that stellar
variability can corrupt any interpretation of a planetary wind.
While \citet{vld+03,vld+08} attempted
to account for this statistically, the out-of-transit time baseline
may be too short to characterize confidently
the intrinsic variability of the stellar Ly$\alpha$ line
(J. Winn, personal communication). There are further problems.
Lyman-$\alpha$ absorption is measurable
only in the line wings because of confusion
near line center from the interstellar medium and from geocoronal (terrestrial)
emission. Even the line wings, however, can be contaminated
by highly time-variable geocoronal emission.

\citet{vdl+04} further used STIS in a lower spectral resolution mode
to measure wavelength-integrated fluxes in various lines. These
measurements enjoyed greater sensitivity to variations in a spectral
line at the cost of not resolving the line profile.  These authors
observed, with 2--3$\sigma$ confidence, the \ion{H}{I} Ly$\alpha$ line
flux to decrease by $\sim$5\% in transit.  Purely from an
observational standpoint, this spectrally unresolved measurement is
claimed to be consistent with the spectrally resolved measurement
\citep{vld+03,vdl+04}. Despite the claimed consistency, the two sets
of measurements are made more than two years apart, which is perhaps
worrisome given how stellar chromospheric emission is highly time
variable. Moreover, it is, of course, impossible to verify whether the
spectrally unresolved observations pertain to the same strikingly
large Doppler equivalent velocities of +/- 100 km/s that are so
clearly implicated in the spectrally resolved observations. In any
case, decrements in \ion{O}{I} and \ion{C}{II} line emission were also
observed, with similarly marginal confidence. Unfortunately, none of
these tantalizing measurements could be reproduced after the STIS
instrument failed in 2004.  \citet{ehren+08} recently attempted
similarly spectrally unresolved measurements with the Hubble Advanced
Camera for Surveys (ACS).  Their detection of \ion{H}{I} absorption in
transit is uncertain but consistent with previous claims.

Even if the STIS observations do signify a planetary outflow,
the mass loss rate implied is not necessarily
large enough to seriously reduce the
mass of the planet. \citet{vld+03} claim a mass loss rate
of $10^{10} \gm \s^{-1}$ based on a fit to the absorption depth at large
Dopper equivalent velocities vs.~time,
and on considerations of how radiation pressure 
can shape the cloud of hydrogen as it expands away from the planet. Note
that radiation pressure is not necessarily claimed by these authors
to drive the outflow; they acknowledge the need for hydrogen
to be heated by the star and to
expand to distances approaching the Roche lobe \citep{ve+04}.\footnote{
In \S\ref{sec-radpress}, we show that radiation pressure
acting on hydrogen through the Ly$\alpha$ transition
adds at most 1\% to the maximum mass loss rates achievable from
a thermal wind heated by photoionization.}
Theoretical models of thermal winds heated by photoionization
generally produce mass loss rates that are
several times $10^{10}\gm \s^{-1}$ \citep{y04,y06,ttp+05,g07}.
Over the several Gyr age of
the system, somewhat less than 1\% of a hot Jupiter's mass
would be carried away by such thermal winds. \citet{g07}
presents particularly convincing hydrodynamic escape models.

Nevertheless, the question remains whether the various STIS observations of HD
209458b are correctly explained as a planetary wind driven by
photoionization.  It is also unclear whether hot Jupiters lose
significant mass early in their
evolution, when the UV luminosities of their host stars are
enormously higher than their main-sequence values.

In this paper, we demonstrate from a first-principles calculation that a hot Jupiter cannot lose a significant fraction of its mass via a planetary outflow driven by UV photoionization at any stage during its lifetime, including the pre-main-sequence phase.  We further show that the spectrally resolved Lyman-$\alpha$ transit observations of HD 209458b  \citep[][]{vld+03,b07} probe velocities too large to reflect the bulk flow of a hot Jupiter wind.  Observations of Lyman-$\alpha$ absorption at Doppler-equivalent velocities of $\pm$100 km/s require either an additional source of high-velocity H atoms \citep[e.g.,][]{hes+08} or an enhancement in the density of neutrals in or around the planetary wind \citep[e.g.,][]{b08}.  We discuss these possibilities in \S\ref{sec-summary}.

Our standard model is
essentially that of a thermal or ``Parker'' wind---a flow accelerated
by gas pressure from subsonic to supersonic velocities through
a critical sonic point \citep{p58}---with the added complication that
the heating is external, from stellar UV irradiation.
Our model includes realistic heating and cooling,
ionization balance, and tidal gravity, but is simple
enough that we can elucidate the basic physics and write down
approximate analytic formulae for the mass loss rate.

\citet{g07} notes correctly that the planetary outflow need not
take the form of a transonic wind.\footnote{In our terminology, ``transonic''
refers to the Parker wind which
transitions from subsonic to mildly supersonic velocities,
while \citet{g07} uses the word to describe a breeze whose peak
velocity is nearly sonic but which eventually decelerates to zero velocity
at infinite distance. For the wind he reserves the word ``supersonic.''}
He points out that the host star wind
can pressure-confine the planetary outflow down to a subsonic breeze.
Stellar wind interactions further complicate the planetary flow 
at large distances \citep[cf.][who model
stellar wind interactions with an entirely neutral wind escaping from
HD 209458b at large velocities]{sve+07}.  By using our hydrodynamics
code to compute breeze solutions, we characterize
qualitatively the extent to which outflows from hot Jupiters
are suppressed, both in the case of a wind emitted by a main-sequence Sun-like
star, and in the case of a pre-main-sequence T Tauri wind.

In \S\ref{sec-model}, after reviewing some of the basic orders
of magnitude characterizing hot Jupiter atmospheres, we
present our standard model of a steady transonic
wind.  We solve the equations of
ionization balance and of mass, momentum, and energy conservation
using a relaxation code, and we demonstrate the robustness of our
solution by exploring a variety of input boundary conditions.
The results of our model are presented for UV fluxes
spanning four orders of magnitude, ranging from those expected for
quiet main-sequence solar analogs to those emitted by active T Tauri stars.  
Section \ref{sec-discuss} contains a wide-ranging discussion
of various aspects of our wind solution. Of especial interest
are how the physics of mass loss changes as the UV flux
increases from low main-sequence values to high pre-main-sequence
values (\S\ref{sec-twoexplain}); how host star winds can squash dayside
planetary outflows, thereby perhaps energizing nightside outflows
(\S\ref{sec-breeze});
and how radiation pressure 
is ineffective compared to UV photoionization heating in
driving an outflow (\S\ref{sec-radpress}).
Finally, \S\ref{sec-summary} summarizes our findings,
pinpoints why the estimate of \citet{lsr+03} and subsequent
determinations by \citet{bsc+04} and \citet{bcb+05} of mass loss rates
are erroneously high, and assesses the STIS observations,
both spectrally resolved and unresolved.

\section{THE MODEL}\label{sec-model}

We construct a simple 1D model
for photoevaporative mass loss from a hot Jupiter.
We focus on the flow originating from the substellar 
point on the planet, and assume that mass loss
occurs in the form of a steady, hydrodynamic, transonic wind.
These restrictions imply that we will calculate
a maximum flux of mass from the planet, insofar as (a) the substellar
point receives the maximum UV flux from the star, (b) tidal gravity
acts most strongly along the substellar ray to accelerate gas away from
the planet, and (c) the transonic wind
carries more mass than pressure-confined, subsonic
breezes.\footnote{For
an introduction to Parker's \citeyearpar{p58} theory of transonic winds
and subsonic breezes, see, e.g.,
``Introduction to Stellar Winds,''
the textbook by \citet{lc99}.}
Applying our solution for the substellar, transonic streamline
over all $4\pi$ steradians yields a hard upper limit on the
total rate of photoevaporative mass loss. How closely
the actual rate of mass loss approaches this upper limit
is discussed in \S\ref{sec-discuss}.

We assume the base of the wind is composed of
atomic hydrogen and calculate how the flow becomes increasingly
ionized as it approaches the star.  We neglect the molecular chemistry
of hydrogen and do not capture the H$_2$/H dissociation front.  This
simplification can be justified by showing that the temperature of the
wind is higher than the $\sim$2000 K required to thermally dissociate
H$_2$, and by showing that above the $\tau = 1$ surface to photoionization,
our solution is insensitive to our chosen boundary conditions.  This we do
in Appendix \ref{sec-bcsense}; a summary is given at the end of
\S\ref{sec-bcs}.  We further neglect helium and metals.
We comment on the implications of this omission in \S\ref{sec-summary}.

The rest of this section is organized as follows.
In \S\ref{sec-eqns} we write down the basic steady-state equations of mass,
momentum, energy, and ionization balance.
The numerical methods used to solve these coupled
ordinary differential equations are detailed
in \S\ref{sec-nummeth}, which includes a listing
of our boundary conditions.
In \S\ref{sec-modres} we present the results of the model.
For those wishing to skip to the punchline,
a simple analytical description of our results may be found
in \S\ref{sec-twoexplain}.

To help orient the reader, we now supply some of our standard model parameters,
together with several order-of-magnitude estimates characterizing the wind.
At Lyman continuum wavelengths, the solar UV luminosity is
roughly $10^{-6}L_{\odot}$, where $L_{\odot}$ is the bolometric solar
luminosity \citep{wrb+98}.
To the extent that host stars of hot Jupiters are like the Sun,
a hot Jupiter with an orbital semi-major axis of $a = 0.05$ AU
receives a UV flux of $F_{\rm UV} = 450 \erg\cm^{-2}\s^{-1}$ between
photon energies of 13.6 eV and 40 eV (this is nearly identical
to the flux employed by \citet{g07} but we derive ours independently
of that study). This flux characterizes
``moderate to low solar activity'' \citep{wrb+98},
and is {\it not} averaged over the planetary surface; for
a discussion of the effects of surface averaging, see \S\ref{sec-daynight}.
We take the planet to have mass $M_{\rm p} = 0.7 M_{\rm J} = 10^{30}\gm$
and a fiducial 1-bar radius $R_{\rm p} \equiv 1.4 R_{\rm J} = 10^{10} \cm$,
where $M_{\rm J}$ and $R_{\rm J}$ are, respectively,
the mass and radius of Jupiter. The planet's surface gravity $g\sim 700$ cm/s$^2$ is approximately the same as $g\sim 10^3$ cm/s$^2$ on Earth.
We take the effective radius of the planet's Roche lobe,
inside of which the planet's gravity dominates the
host star's tidal gravity, to equal the approximate distance
to the planet's L1 point:
$R_{\rm Roche} = [M_{\rm p}/(3M_{\ast})]^{1/3}a = 4.5 R_{\rm p}$,
where $M_{\ast} = M_{\odot}$ is the mass of the star.
In many astrophysical
situations---including ours, as will be shown---photoionized gas
cools by radiation from collisionally excited atomic hydrogen, which
thermostats the gas temperature $T$ to $\sim$$10^4\K$.
The corresponding sound speed is $\sim$10 km/s.
The hydrostatic pressure scale height is $H = kT/(m_{\rm H} g)
\sim 0.1 R_{\rm p}$,
where $k$ is the Boltzmann constant and $m_{\rm H}$ is the mass of
the hydrogen atom (of course the wind is not strictly hydrostatic
but it is nearly so near its base where speeds are still subsonic).
To travel a distance $R_{\rm p}$ at the sound
speed takes a few hours.

Finally, we can estimate the gas density
and pressure where the wind is launched, i.e.,
where the bulk of the stellar UV radiation is absorbed.
At a photon energy of $h\nu_0 = 20$ eV, the cross section for photoionization
of hydrogen
is $\sigma_{\nu_0} = 6 \times10^{-18} (h \nu_0/13.6{\rm\,eV})^{-3}\cm^2$
\citep[e.g.,][]{s78}; optical depth unity is achieved in a neutral column
$N_{\rm H} = 1/\sigma_{\nu_0} = 5 \times 10^{17} \cm^{-2}$;
dividing this column by the scale height $H$ gives a neutral density
$n_0 \sim 6 \times 10^8 \cm^{-3}$ (equivalently, a neutral mass
density $\rho \sim 10^{-15}\gm \cm^{-3}$); and multiplying by $kT$ gives
a partial pressure $P \sim 1$ nanobar at the base of the wind.
By contrast, visible radiation from
the star is absorbed at pressures closer to 1 bar, setting the temperature
below the base of the wind to be
$T_{\rm below} \sim 10^3$ K and the pressure scale height to be
$H_{\rm below} = kT_{\rm below}/(2m_{\rm H}g) \sim 0.005 R_{\rm p}$, where
the factor of 2 accounts for the fact that the hydrogen at depth is molecular.
The smallness of $H_{\rm below}$ means that the wind is launched
at a radius very nearly equal to $R_{\rm p}$:
reducing the pressure from 1 bar to 1 nanobar takes
about 20 scale heights or 0.1$R_{\rm p}$. In other words,
the radius at which UV photons are absorbed is approximately
$1.1R_{\rm p}$. This radius enters significantly into the magnitude
of the mass loss rate, as we discuss in \S\ref{sec-twoexplain},
\S\ref{sec-summary}, and Appendix \ref{sec-bcsense}.

\subsection{Basic Equations}\label{sec-eqns}
As stated above, we concentrate on the streamline originating from the
substellar point on the planet. From mass continuity,
\begin{equation}\label{eqn-con}
\frac{\partial}{\partial r}\left(r^2\rho v\right) = 0 \,\, ,
\end{equation}
where $r$ is the distance from the center of the planet to the star,
and the gas has density $\rho$ and velocity $v$.  In the frame rotating with the
planet's orbital frequency, momentum conservation implies
\begin{equation}\label{eqn-mom}
  \rho v \frac{\partial v}{\partial r} = -\frac{\partial P}{\partial r} - \frac{GM_{\rm p}\rho}{r^2} + \frac{3GM_*\rho r}{a^3} \,\, ,
\end{equation}
where 
$G$ is the gravitational constant.  We call the last term on the
right-hand side of (\ref{eqn-mom}) the ``tidal gravity'' term: it
is the sum of the centrifugal force and differential stellar gravity,
along the ray joining the planet to the star (neglecting the small
shift in the system barycenter away from the star; cf.~\citealt{g07}).
For simplicity, we neglect
the Coriolis force, the magnitude of which is
comparable to the magnitude of other forces only at the outer boundary
of our calculation, near the Roche lobe radius where gas moves at approximately
the planet's escape velocity.

The equation for energy conservation is
\begin{equation}\label{eqn-energy}
 \rho v \frac{\partial}{\partial r}\left[\frac{kT}{(\gamma-1)\mu}\right] =  \frac{kTv}{\mu}\frac{\partial\rho}{\partial r}  + \Gamma + \Lambda \,\, ,
\end{equation}
where the left-hand side tracks changes in the internal thermal energy
of the fluid, $\mu$ is the mean molecular weight,
and $\gamma = 5/3$ is the usual ratio of specific heats
for a monatomic ideal gas. 
On the right-hand side we have three terms
denoting, respectively, cooling due to $PdV$ work done by expanding
gas, heating from photoionization, and cooling from radiation and
conduction. We do not include a term proportional to 
the chemical potential because changes in energy due to changes
in the number of particles are already accounted for
in our photoionization term $\Gamma$.
Equation (\ref{eqn-energy}) follows in a straightforward
way from the standard steady-state energy equation,
$\nabla \cdot (\rho u \mathbf{v}) = - P \nabla \cdot \mathbf{v} + \Gamma +
\Lambda$, after using $\rho v r^2$ = constant and $u = kT/[(\gamma-1)\mu]$
for the specific internal energy.

We assume for simplicity
that the UV flux is concentrated
at one photon energy $h\nu_0 = 20$ eV. Then
\begin{equation}\label{eqn-Q0}
\Gamma = \varepsilon F_{\rm UV}e^{-\tau} \sigma_{\nu_0} n_0 \,\, ,
\end{equation} 
where $n_0$ is the number density of neutral H atoms,
$\varepsilon = (h\nu_0 - 13.6{\rm\,eV})/h\nu_0$ is
the fraction of photon energy deposited as heat,\footnote{This
is a maximum efficiency and can be somewhat smaller if there are
other ways for the primary photoelectron to deposit its
energy \citep[e.g., by secondary ionization; cf.][]{wck+83}.}
and 
\begin{equation}\label{eqn-tau}
\tau = \sigma_{\nu_0}\int_r^\infty n_0\,dr
\end{equation}
is the optical depth to ionizing photons.

Equation (\ref{eqn-Q0}) assumes that photoelectrons share their
kinetic energy with other gas species locally. We have verified
{\it a posteriori} that this assumption
is justified. For our standard model, at the wind base,
a photoelectron travels $\lambda_{\rm mfp} \sim 2 \times 10^{-4}R_{\rm p}$
before colliding with a neutral
H atom, assuming a cross section of $\sim$$10^{-15}\cm^{2}$.
The number of collisions required for the electron to give up most
of its energy to surrounding H atoms is $\sim$$m_{\rm H}/m_{\rm e}$, where
$m_{\rm e}$ is the electron mass; the corresponding distance random walked
is $(m_{\rm H}/m_{\rm e})^{1/2}\lambda_{\rm mfp} \sim 7 \times 10^{-3} R_{\rm p}$.
A similarly short distance obtains
at the outer periphery of our calculation, near the Roche lobe,
where photoelectrons share their energy with H$^+$ ions via
Coulomb collisions. The margin of safety is still larger
for our high flux model (\S\ref{sec-highflux}),
for which ion densities are greater.

The main contribution to $\Lambda$ is Ly$\alpha$ radiation, emitted by
neutral H atoms that are collisionally excited by electrons:
\begin{equation}\label{eqn-lyalpha}
\Lambda \approx \Lambda_{{\rm Ly}\alpha} = -7.5\times 10^{-19} n_+n_0 e^{-118348\K/T} \erg \cm^{-3} \s^{-1}
\end{equation}
where $n_+$ is the number density of H$^+$ (= the number
density of electrons), and all densities are
measured in ${\rm cm}^{-3}$ \citep{b81}.
Other cooling mechanisms---collisional ionization, radiative
recombination, free-free emission, and thermal conduction---are
negligible, as shown in \S\ref{sec-modres} (see also Appendix
\ref{sec-other}).
Our assumption that Ly$\alpha$ photons are able to escape and thereby
cool the flow is validated in Appendix \ref{sec-escape}.

In ionization equilibrium, the rate of photoionizations
balances the rate of radiative recombinations plus the rate at which
ions are advected away:
\begin{equation}\label{eqn-ion}
n_0 \frac{F_{\rm UV}e^{-\tau}}{h\nu_0}\sigma_{\nu_0} = n_+^2 \alpha_{\rm rec} + \frac{1}{r^2}\frac{\partial}{\partial r}\left(r^2 n_+ v\right) \,\, ,
\end{equation}
where $\alpha_{\rm rec} = 2.7\times 10^{-13}(T/10^4\K)^{-0.9}$
is the Case B radiative recombination
coefficient for hydrogen ions \citep{sh95}.
By continuity, the advection term can be rewritten as
\begin{equation}
\frac{1}{r^2}\frac{\partial}{\partial r}\left(r^2 n_+ v\right) = nv\frac{\partial f_+}{\partial r} \,\, ,
\end{equation}
where the ionization fraction $f_+ = n_+/n$ and $n = n_+ + n_0$
is the total number density of hydrogen nuclei.
Note that $\mu = m_{\rm H}/(1+f_+)$.
Collisional ionization is negligible compared to photoionization, as
demonstrated in \S\ref{sec-modres}.

\subsection{Numerical Method}\label{sec-nummeth}

The problem of finding the structure of the wind is a two point boundary
value problem \citep[for an introduction into the nature of such problems, see, e.g.,][]{ptv+92}. The two points are
the base of the flow and the sonic point.
We have solved the problem by constructing a relaxation code.
In our case relaxation methods are preferred over shooting methods
because for every transonic wind solution there are an infinite
number of breeze solutions \citep{p58}. Furthermore,
the sonic point is a critical point where derivatives, if not carefully
computed, can become singular.
Instead of searching exhaustively in a multidimensional
space for the one solution that ``threads the needle'' of the
critical sonic point, it is more efficient to start
with an approximate solution that already satisfies
the sonic point conditions, and refine that solution
to higher accuracy. 
Previous attempts that did not use
relaxation algorithms to find transonic winds
found, not surprisingly,
breezes instead \citep{kp83,y04}.
Relaxation methods are also suitable
for our problem because the wind profile is expected to be smooth,
with no oscillatory behavior, and so pre-defining a radial grid
for the solution is not especially problematic.
Nevertheless, care needs to be taken in implementing the 
method; we describe as follows our procedure, developed
after considerable experimentation. Sections \ref{sec-fde}---\ref{sec-complicate} describe how we compute the wind profile from some base depth
in the atmosphere up to the sonic point; section \ref{sec-hill} takes
this solution and extends it out to the Roche lobe.

For alternate numerical methods, see \citet{ttp+05} and \citet{g07}.

\subsubsection{Finite Difference Equations: From the Base to the Sonic Point}\label{sec-fde}

From a location $r_{\rm min} = R_{\rm p}$
in the upper atmosphere of the planet to the
sonic point $r_{\rm s}$ of the wind, we use the Numerical Recipes
relaxation routine \texttt{solvede} \citep{ptv+92} to solve
the finite difference versions of (\ref{eqn-con}),
(\ref{eqn-mom}), (\ref{eqn-energy}), (\ref{eqn-tau}), and (\ref{eqn-ion}):
\begin{eqnarray} \label{eqn-rho}
E_{1j} &\equiv& \Delta_j \rho -  \frac{d\rho}{dr} \Delta_j r \nonumber \\
&=& \Delta_j \rho + \rho\left(\frac{2}{r} + \frac{1}{v}\frac{dv}{dr} \right)\Delta_j r = 0 \\
\label{eqn-vel}
E_{2j} &\equiv& \Delta_j v -  \frac{dv}{dr}\Delta_j r \nonumber \\
&=& \Delta_jv - \frac{v}{v^2-\gamma kT/\mu}\bigg[2\gamma kT/(\mu r) - (\gamma-1)Q/(\rho v) \nonumber \\
&& \;\;\;\;\;\;\;\;\; - GM_{\rm p}/r^2 + 3GM_*r/a^3\bigg] \Delta_j r = 0 \\
\label{eqn-T}
E_{3j} &\equiv& \Delta_jT - \frac{dT}{dr} \Delta_j r \nonumber \\
&=& \Delta_j T -  \left[(\gamma -1)\left(\frac{Q}{\rho v}\frac{\mu}{k} + \frac{T}{\rho}\frac{d\rho}{dr}\right)\right. \nonumber \\
&& \;\;\;\;\;\;\;\;\; \left.- \frac{T}{(1+f_+)}\frac{df_+}{dr}\right]\Delta_j r = 0 \\
E_{4j} &\equiv& \Delta_j \tau - \frac{d\tau}{dr} \Delta_jr \nonumber \\
&=& \Delta_j\tau + \frac{(1-f_+)\rho}{m_{\rm H}} \sigma_{\nu_0} \Delta_j r = 0 \\
\label{eqn-ionbal}
E_{5j} &\equiv& \Delta_j f_+  - \frac{df_+}{dr} \Delta_jr \nonumber \\ 
&=& \Delta_jf_+ - \frac{m_{\rm H}}{\rho v}\left[ \frac{F_{\rm UV}e^{-\tau}}{h\nu_0}\sigma_{\nu_0}\frac{(1-f_+)\rho}{m_{\rm H}} \right. \nonumber \\
&& \;\;\;\;\;\;\;\;\; \left.- \alpha_{\rm rec} \left(\frac{f_+\rho}{m_{\rm H}}\right)^2 \right]\Delta_j r = 0
\end{eqnarray}
where $Q\equiv \Gamma+\Lambda$,
and $\Delta_jx = x_j - x_{j-1}$ at the $j$th radial grid point.
In evaluating the individual variables that make up the derivatives, we
average across adjacent grid points; e.g.,
$\rho = (\rho_j+\rho_{j-1})/2$. This is the same choice
adopted by \citet{ptv+92}.

We introduce
an extra dependent variable $z \equiv r_{\rm s} - r_{\rm min}$
because we do not know {\it a priori}
the location of the sonic point $r_{\rm s}$.  Thus
\begin{equation}\label{eqn-z}
E_{6j} \equiv \Delta_j z \equiv \Delta_j (r_{\rm s} - r_{\rm min}) = 0 \,\,.
\end{equation}
In other words, we solve for $z$ just like we do any other dependent variable,
and its solution tells us the radial location of the sonic point ($r_{\rm s}$).

Our six dependent variables are $\rho$, $v$, $T$, $f_+$, $\tau$, and $z$, all
non-dimensionalized for ease of calculation
using the scales
$\rho_0 = 10^{-15}$ g/cm$^3$, $T_0 = 10^4$ K, $v_0 = (k T_0/m_{\rm H})^{1/2} = 9$ km/s,
and $z_0 = 10^{10}$ cm.
We solve for these variables on a radial grid of 1000 points
that has more points concentrated near
$r_{\rm min}$ (where derivatives are large) and near $r_{\rm s}$.
The convergence parameter ``conv'' that \texttt{solvede} uses
is set to $10^{-10}$.
The scale parameters
``scalv'' used to calculate the convergence parameter are $\rho = 100$,
$v=2$, $T=1$, $\tau=100$, $f_+=1$, and $z=3$ in our non-dimensionalized units. 
We take
our independent, radius-like variable to be $q$ such that
$r = r_{\rm min}+qz$; $q$ runs from 0 to 1. 

The finite difference equations are solved
by a multidimensional Newton's method, which requires
that we evaluate partial derivatives of the $E_{ij}$'s
with respect to the dependent variables. We evaluate
these partial derivatives numerically, by introducing
small finite changes in each of the dependent variables
and calculating the appropriate differences.

\subsubsection{Boundary Conditions}\label{sec-bcs}

We need six boundary conditions (BC) to solve Equations (\ref{eqn-rho})--(\ref{eqn-z}).  
Two boundary conditions are provided by the requirement that the wind
pass through the critical point $r_{\rm s}$ where
\begin{equation}\label{eqn-bc1}
\left[v^2 = \frac{\gamma kT}{\mu}\right]_{r_{\rm s}} \,\,\,\,\,\, ({\rm BC1})
\end{equation}
and
\begin{equation}\label{eqn-bc2}
\left[\frac{2\gamma kT}{\mu} - \frac{GM_{\rm p}}{r}  - \frac{(\gamma-1)Qr}{\rho v} + \frac{3GM_{\ast}r^2}{a^3}\right]_{r_{\rm s}} = 0 \,\,\,\,\,\, ({\rm BC2})
\end{equation}
in order to avoid infinite derivatives (see Equation \ref{eqn-vel}).
These are the critical point conditions of the \citet{p58} transonic wind
\citep[see, e.g.,][]{lc99}.

We choose our remaining four boundary conditions as follows.
At the base of the flow, we set
$\rho(r_{\rm min}) = 4 \times 10^{-13} \gm \cm^{-3}$ (BC3); 
$f_+(r_{\rm min}) = 10^{-5}$ (BC4); 
and $T = 1000$ K (BC5),
approximately the effective temperature
of a hot Jupiter at $a = 0.05$ AU \citep[e.g.,][]{bsh03}.
For our final boundary condition, we enforce the condition
that $\tau(r_{\rm s})$ equals the optical depth between
the sonic point and the Roche lobe---see \S\ref{sec-hill}.
For our standard model, this optical depth turns out to be
$\tau(r_{\rm s}) = 0.0023$ (BC6).

In Appendix \ref{sec-bcsense}, we demonstrate that our solution
is insensitive to these particular choices of numbers for
BC3 through BC6.
We show there that the solution hardly changes
as long as $\rho(r_{\rm min})$ is large enough
that $\tau(r_{\rm min}) \gg 1$; $f_+(r_{\rm min})\ll 1$;
$T(r_{\rm min}) \ll 10^4$ K; and $\tau(r_{\rm s}) \ll 1$.
We also describe how the mass loss rate changes by less than a factor of
2 for 10\% variations in $r_{\rm min}$ about our fiducial
radius $R_{\rm p} = 10^{10} \cm$ (see the order-of-magnitude
discussion at the beginning of \S\ref{sec-model}
for why $r_{\rm min}$ is only uncertain by about 10\%).
The insensitivity to boundary conditions helps to justify
our neglect of the H$_2$/H dissociation front, which is located at greater
depth than
the H/H$^+$ ionization front---the latter we do resolve, near $r_{\rm min}$.

\subsubsection{Sonic Point Limit}\label{sec-expand}
Because the exact expression for
$dv/dr$ in Equation (\ref{eqn-vel})---and, by extension,
Equations (\ref{eqn-rho}) and (\ref{eqn-T})---is
difficult to evaluate accurately near the sonic
point (both numerator and denominator of $dv/dr$ vanish there),
we have derived an analytic form for it that
is strictly valid only at the sonic point:
\begin{eqnarray}\label{eqn-expand}
\left.\frac{dv}{dr}\right|_{r_{\rm s}} &=& \frac{\gamma-1}{\gamma+1}\left\{-\frac{2v}{r} + \frac{\gamma Q}{2\rho v^2} - \frac{Q_1}{2\rho v} \right.\nonumber \\
 &+& \frac{1}{2v}\left[\left(\frac{\gamma Q}{\rho v}\right)^2 + \frac{8v^2Q_1}{\rho r}-\frac{2\gamma QQ_1}{\rho^2v} + \frac{Q_1^2}{\rho^2} \right. \nonumber \\
 &-& 4\frac{(\gamma+1)}{(\gamma-1)}\frac{vQ_2}{\rho} - \frac{16}{(\gamma-1)}\frac{Qv}{\rho r} + \frac{8(5-3\gamma)}{(\gamma -1)^2}\frac{v^4}{r^2} \nonumber \\
 &+& \left.\left.\frac{36(\gamma+1)v^2}{(\gamma-1)^2}\frac{GM_*}{a^3}\right]^{1/2}\right\}
\end{eqnarray}
where $Q_1$ and $Q_2$ are defined by $dQ/dr \equiv Q_1 (dv/dr) + Q_2$.
This expression is derived by applying L'H\^{o}pital's rule to
$dv/dr$ in Equation \ref{eqn-vel}, and simplifying the result
using boundary conditions (\ref{eqn-bc1}) and (\ref{eqn-bc2}).

In our code, we evaluate $dv/dr$ as
\begin{equation}
\frac{dv}{dr} = F_{\rm exact} \left.\frac{dv}{dr}\right|_{\rm exact} + (1-F_{\rm exact}) \left.\frac{dv}{dr}\right|_{r_{\rm s}}
\end{equation}
where
$$
F_{\rm exact} = -{\rm erf}\left[p\left(1- \frac{\gamma kT}{\mu v^2}\right)\right] \,\,,
$$
${\rm erf}$ is the error function, $p=100$ is a parameter that determines 
the width of the transition between the two right-hand terms,
and $(dv/dr)_{\rm exact}$ is given by Equation (\ref{eqn-vel}).
Far from the sonic point, $(dv/dr)_{r_{\rm s}}$ is not accurate,
so where $F_{\rm exact}$ is within machine precision of 1, we revert
to using $(dv/dr)_{\rm exact}$ only.

\subsubsection{Solving Successively More Complicated Problems}\label{sec-complicate}
Relaxation codes require good initial guesses to converge,
so we build our final solution by solving successively
more complicated problems. The solution of a given problem
furnishes the initial guess for the next problem.

First we use our relaxation code to find
solutions for the separate problems of an isothermal
wind without photoionization, and an isothermal hydrostatic atmosphere with
photoionization, both neglecting tidal gravity.
The combination of these solutions provides the
initial guess for an isothermal wind with photoionization.
Then we remove the restriction that the wind be isothermal.
The photoionization heating term, followed by cooling terms
and finally the tidal gravity term,
are added one by one. Sometimes the addition
of even a single term requires iteration: the term
must be added in diluted form first and gradually
strengthened to full amplitude.

\subsubsection{From the Sonic Point to the Roche Lobe}\label{sec-hill}
From the sonic point outward,
we use the Numerical Recipes routine \texttt{odeint} with a Bulirsch-Stoer integrator \citep{ptv+92} to solve our original ordinary
differential equations (\S\ref{sec-eqns}).
We set the routine's convergence parameter ``eps'' to $10^{-13}$.
The solution is extended out to the planet's Roche lobe radius
$R_{\rm Roche}$.
This extension is a straightforward initial value problem,
with initial conditions at the sonic point provided by our relaxation solution.
For the first few radial steps in the integration, we use 
(\ref{eqn-expand}) for $dv/dr$ to avoid numerical singularities.

Finally, this extended solution feeds back into our relaxation code
through $\tau(r_{\rm s})$. We iterate a few times between the relaxation
code and the Bulirsch-Stoer integrator until $\tau(r_{\rm s})$
self-consistently reflects
the optical depth between the sonic point and the Roche lobe radius.
The assumption here is that the optical depth beyond the Roche lobe
is negligible; this assumption is valid because $\tau(r_{\rm s}) \ll 1$
for solutions of interest to us (see Appendix \ref{sec-bcsense}).

\subsection{Results}\label{sec-modres}

In \S\ref{sec-standardres},
we present the results for our standard model, appropriate for
a hot Jupiter orbiting a Sun-like star on the main sequence.
In \S\ref{sec-highflux}, we present a sample high flux case for which
$F_{\rm UV}$ is increased a thousandfold over its standard value, as would
be the case for a hot Jupiter orbiting an active pre-main-sequence star.
In \S\ref{sec-two}, we describe how the maximum mass loss
rate $\dot{M}$ varies when $F_{\rm UV}$ ranges over four decades.

\subsubsection{Standard Model: Hot Jupiter Orbiting Main-Sequence Star}\label{sec-standardres}
\begin{figure}
\centering
\scalebox{1.2}{\plotone{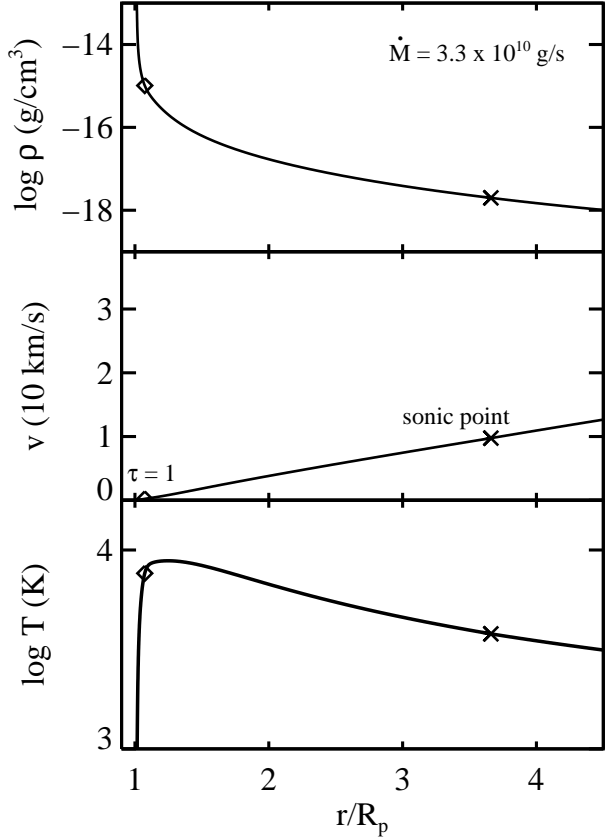}} 
\caption{Standard planetary wind model, which uses parameters
inspired by the hot Jupiter HD 209458b.
A UV flux of $450 \erg \cm^{-2} \s^{-1}$ incident on a planet with mass
$0.7M_{\rm J}$ and radius $R_{\rm p} = 1.4R_{\rm J}$
located 0.05 AU from a $1M_\sun$ star drives a transonic wind.
The flow is calculated along the line joining the planet to the star.
Density $\rho$ (top), wind speed $v$ (middle), and temperature $T$ (bottom) are presented as functions of altitude.
On each panel, the sonic point of the wind is marked with an x,
and the $\tau = 1$ surface to photoionization is marked with a diamond.
The planet loses mass at a maximum rate of $\dot M = 4\pi r^2\rho v = 3.3\times 10^{10}\,\rm{g/s}$.}
\label{fig-450rhovT}
\end{figure}

\begin{figure}
\centering
\scalebox{1.2}{\plotone{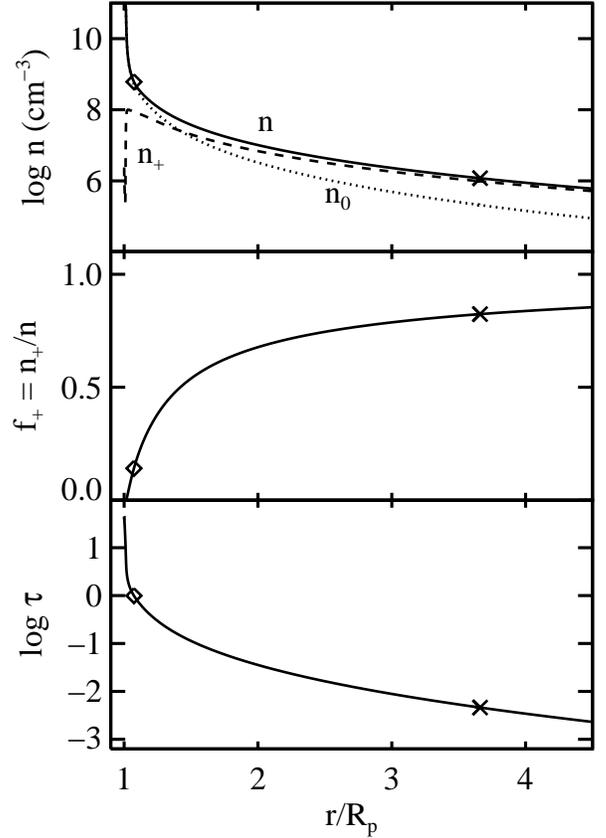}} 
\caption{Planetary wind model for the same standard parameters
  presented in Figure \ref{fig-450rhovT}. Number densities of neutral
  plus ionized hydrogen $n$, neutral hydrogen $n_0$, and ionized hydrogen $n_+$ (top);
  ionization fraction $f_+ = n_+/n$ (middle); and optical depth $\tau$
  to photoionization (bottom) are presented as functions of
  altitude. On each panel, the sonic point of the wind is marked with
  an x, and the $\tau = 1$ surface is marked with a diamond. As much
  as $\sim$20\% of the hydrogen remains neutral at high altitude.}
\label{fig-450nfptau}
\end{figure}

\begin{figure}
\centering
\scalebox{1.2}{\plotone{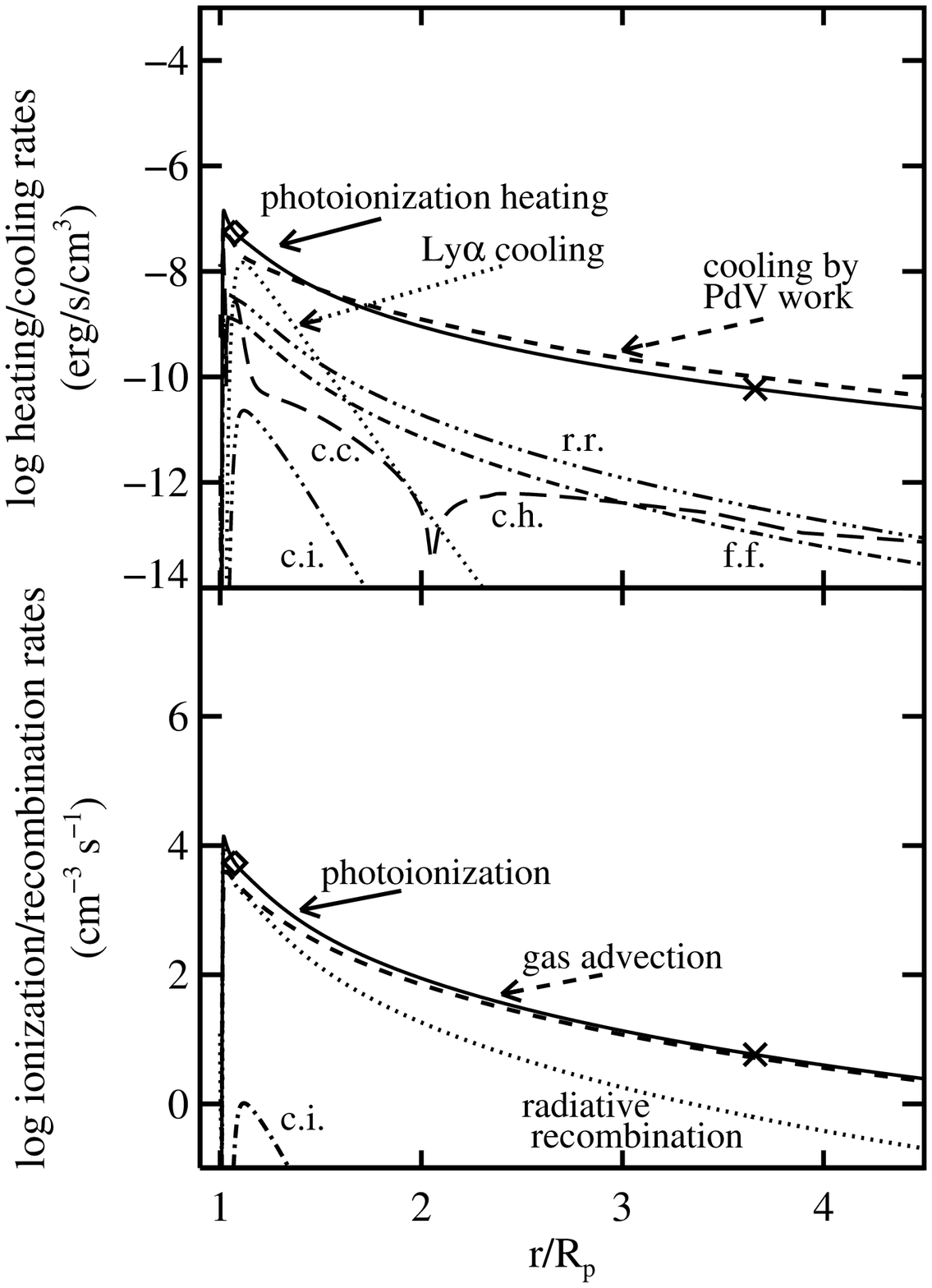}} 
\caption{Contributions to energy equilibrium (top)
and ionization equilibrium (bottom)
for the standard transonic wind 
displayed in Figures \ref{fig-450rhovT} and \ref{fig-450nfptau}.   
In each panel, the sonic point of the wind is marked with an x,
and the $\tau = 1$ surface to photoionization is marked with a diamond.
Energy balance is dominated by photoionization heating and $PdV$ work
done by expanding gas, with some contribution from
Ly$\alpha$ cooling near the base of the wind.
Internal energy changes due to recombination radiation (r.r.),
collisional ionization (c.i.),
free-free radiation (f.f.), and conduction (either cooling c.c. or heating c.h.)
are negligible (see Appendix A for the formulae used to evaluate
these extra contributions).
UV photoionization is balanced by gas advection above the $\tau = 1$ surface.
}
\label{fig-450terms}
\end{figure}

Figures \ref{fig-450rhovT}--\ref{fig-450terms} display the results for
our standard model. This numerical solution verifies many of the
order-of-magnitude estimates made at the beginning of \S\ref{sec-model}.
The stellar UV flux drives a transonic wind with temperature
$\sim$$10^4$ K and velocity $\sim$10 km/s.
The hydrogen density where optical depth unity to Lyman continuum photons
is reached is $\sim$$10^9 \cm^{-3}$.
Because our solution pertains to the substellar
ray connecting the planet to the star,
it yields the maximum mass flux $\rho v$; escape is most
aided by tidal gravity along this ray,
and the substellar point receives the greatest UV flux.
Applying this solution over the entire surface of
the planet yields an upper bound on the mass loss rate of $\dot M = 3.3\times
10^{10}$ g/s.  In \S\ref{sec-daynight}--\ref{sec-breeze},
we estimate the factors
by which the actual mass loss rate is reduced.

Figure \ref{fig-450terms} (bottom panel)
displays the relative contributions to ionization
balance as a function of altitude.  Above the $\tau=1$ surface, gas
advection, not radiative recombination, balances UV photoionization.
A hydrogen atom, once photoionized, does not have time to recombine
before it is swept outward with the wind.
As a gas parcel travels outward, its electron density decreases and
the recombination time $1/(n_+\alpha_{\rm rec})$ becomes ever longer;
more and more of its atoms
are ionized, and the ionization fraction increases with altitude.
At the sonic point, about 20\% of the hydrogen remains neutral.  
This situation differs from static \ion{H}{II} regions
in which photoionization is balanced
by radiative recombination and the transition between ionized and
neutral gas is sharp.

Figure \ref{fig-450terms} (top panel) also shows that the temperature
of the gas is largely controlled by heating from photoionizations and
by cooling from gas expansion ($PdV$ work), with a small contribution
from cooling by Ly$\alpha$ radiation at the wind base.  In the upper
portions of the wind, cooling by $PdV$ work lowers the gas temperature
from its peak of 10000 K to about 3000 K (Figure \ref{fig-450rhovT}).
Since these temperatures exceed the $\sim$2000 K required for
H$_2$ to dissociate, our neglect of H$_2$, and by extension the
radiative coolant H$_3^+$ that is known to be important
where H$_2$ exists in abundance \citep{y04,g07}, is self-consistent,
at least above the $\tau=1$ surface to photoionization of atomic H.
In other words, most of our modeled region is self-consistently devoid
of molecular hydrogen.

Integrated over the entire radial extent of our model from $r_{\rm
  min}$ to $R_{\rm Roche}$, a measure of the heating rate per solid
angle (measured from the planet) from photoionizations is $L_{\rm
  photo} \equiv \int \Gamma r^2 dr = 2.3 \times 10^{22} \erg \s^{-1}
\sr^{-1}$, about equal in magnitude to the analogously defined cooling
rate due to $PdV$ work, $L_{PdV} = -1.9 \times 10^{22} \erg \s^{-1}
\sr^{-1}$. In comparison, the height-integrated cooling rate from
Ly$\alpha$ radiation is only $L_{{\rm Ly}\alpha} = -2.9 \times 10^{21}
\erg \s^{-1}\sr^{-1}$.  (Note that $L_{\rm photo} + L_{PdV} + L_{{\rm
    Ly}\alpha} \neq 0$ because the internal energy of the gas changes
along the flow; see Equation \ref{eqn-energy}).  The fact that as much
as 83\% of photoionization heating goes into $PdV$ work implies that
mass loss is largely ``energy-limited''---see \S\ref{sec-two} and
\S\ref{sec-twoexplain}.

We have used our solution to estimate, ex post facto,
the contributions to internal energy changes from
recombination radiation,
free-free emission, collisional ionizations, and conduction.
As demonstrated in Figure \ref{fig-450terms}, these are not significant.

\subsubsection{High UV Flux Case: Hot Jupiter Orbiting a T Tauri Star}\label{sec-highflux}

\begin{figure}
\centering
\scalebox{1.2}{\plotone{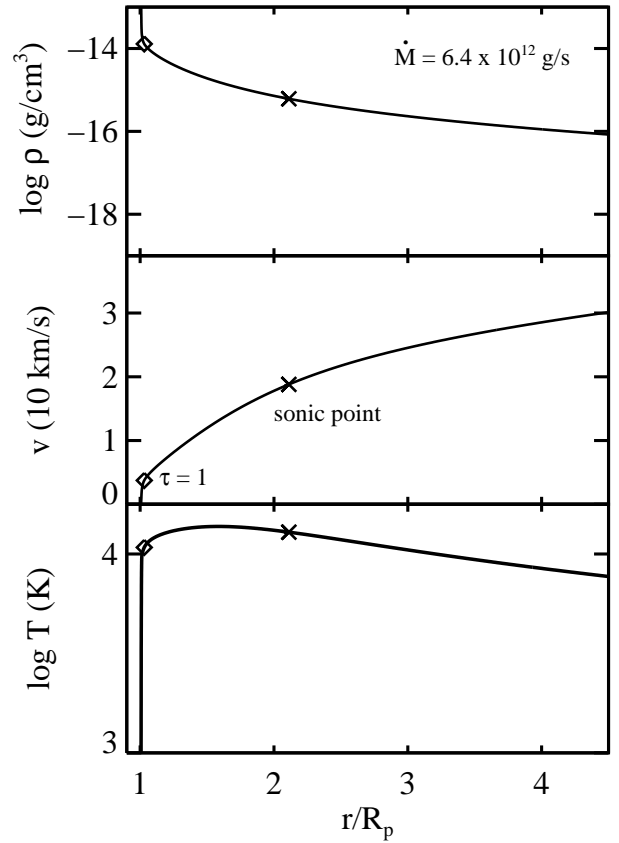}} 
\caption{Same as Figure \ref{fig-450rhovT}, except 
that $F_{\rm UV} =5\times 10^5 \erg \cm^{-2} \s^{-1}$.
Such a flux is characteristic
of that incident on a hot Jupiter orbiting an active pre-main-sequence star.
The planet loses mass at a maximum rate of $\dot M = 4\pi r^2\rho v = 6.4\times
  10^{12}\,\rm{g/s}$. }
\label{fig-5e5rhovT}
\end{figure}

\begin{figure}
\centering
\scalebox{1.2}{\plotone{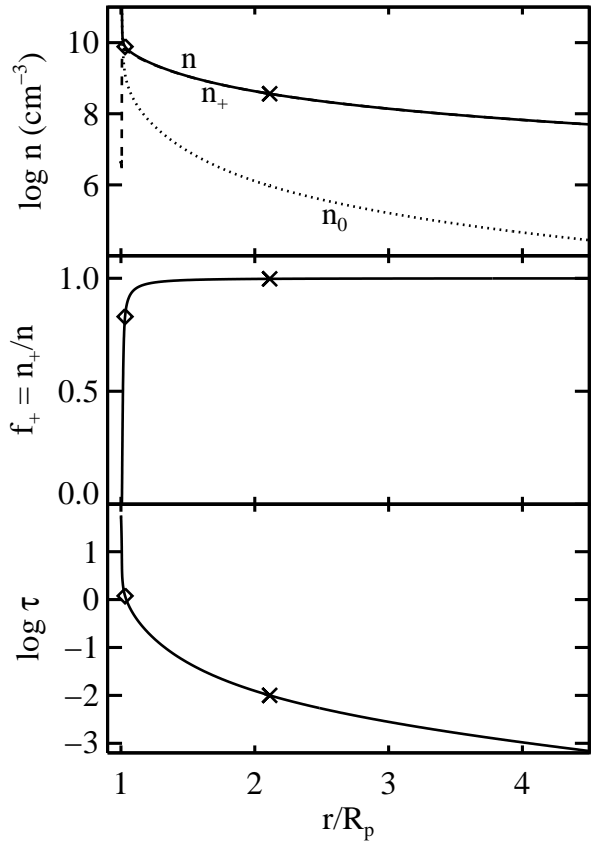}} 
\caption{Same as Figure \ref{fig-450nfptau}, except
for the high UV flux case.
By contrast to the standard model, the wind here is more fully ionized,
and transitions from neutral to ionized more sharply.}
\label{fig-5e5nfptau}
\end{figure}

\begin{figure}
\centering
\scalebox{1.2}{\plotone{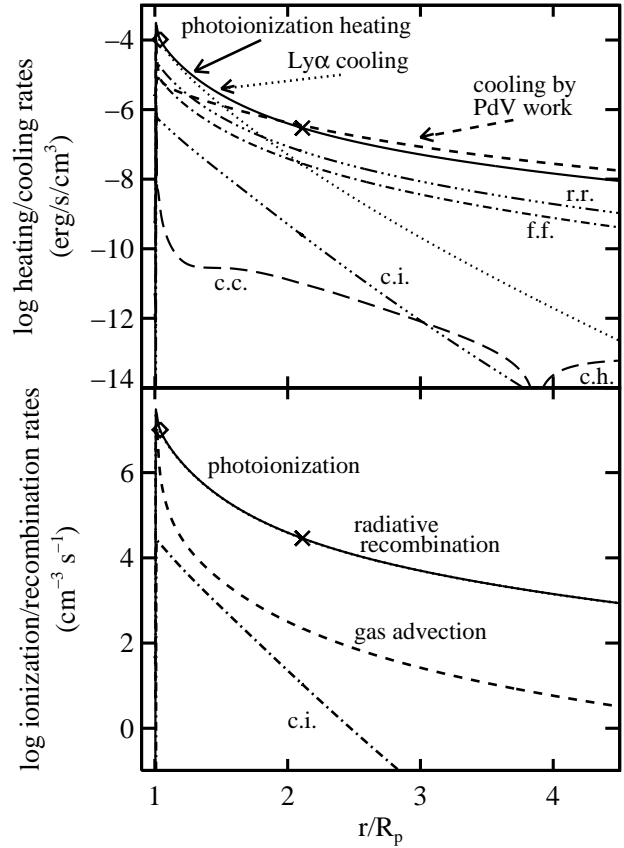}} 
\caption{Same as Figure \ref{fig-450terms}, but for the high
UV flux case. By contrast to the standard model, Ly$\alpha$ cooling
balances photoionization heating at the wind base, and radiative
recombinations balance photoionizations everywhere. The wind here behaves
more nearly like an standard \ion{H}{II} region. 
}
\label{fig-5e5terms}
\end{figure}

Figures \ref{fig-5e5rhovT}--\ref{fig-5e5terms} display
results for a much larger UV flux of
$F_{\rm UV} = 5\times 10^5 \erg \cm^{-2} \s^{-1}$, characteristic
of the radiation field experienced by a hot Jupiter
orbiting a T Tauri star \citep[e.g.,][]{hyj00}.
Again, the UV flux drives a transonic wind with temperature
$\sim$$10^4$ K and velocity a few $\times$ 10 km/s.
The density of the wind is substantially greater, however,
than in the standard model.
Applying our solution over the entire surface
of the planet yields a maximum mass loss rate of $\dot M = 6.4\times
10^{12}$ g/s.  

In contrast to the standard model, here
radiative recombination, not gas advection, balances
photoionization above the $\tau = 1$ surface. 
The transition from a neutral to a nearly fully ionized flow is sharp.
Also in contrast to the standard model, Ly$\alpha$ cooling
plays a dominant role in balancing photoionization heating near
the base of the wind. 
The global energy budget now divides as follows:
$L_{\rm photo}:L_{PdV}:L_{{\rm Ly}\alpha} = 1:-0.29:-0.67$ (again,
$L_{\rm photo} + L_{PdV} + L_{{\rm Ly}\alpha} \neq 0$ because
the internal energy of the gas changes along the flow).
Thus the high flux case is strongly radiatively
cooled. These properties of the high flux case
are similar to those of classic \ion{H}{II} regions
(Str\"omgren spheres). Mass loss under these conditions
is ``radiation/recombination-limited''---see \S\ref{sec-two}
and \S\ref{sec-twoexplain}.

\subsubsection{Mass Loss Rate vs.~UV Flux}\label{sec-two}
\begin{figure}
\centering
\hbox{\hspace{-0.2in}\scalebox{1.3}{\plotone{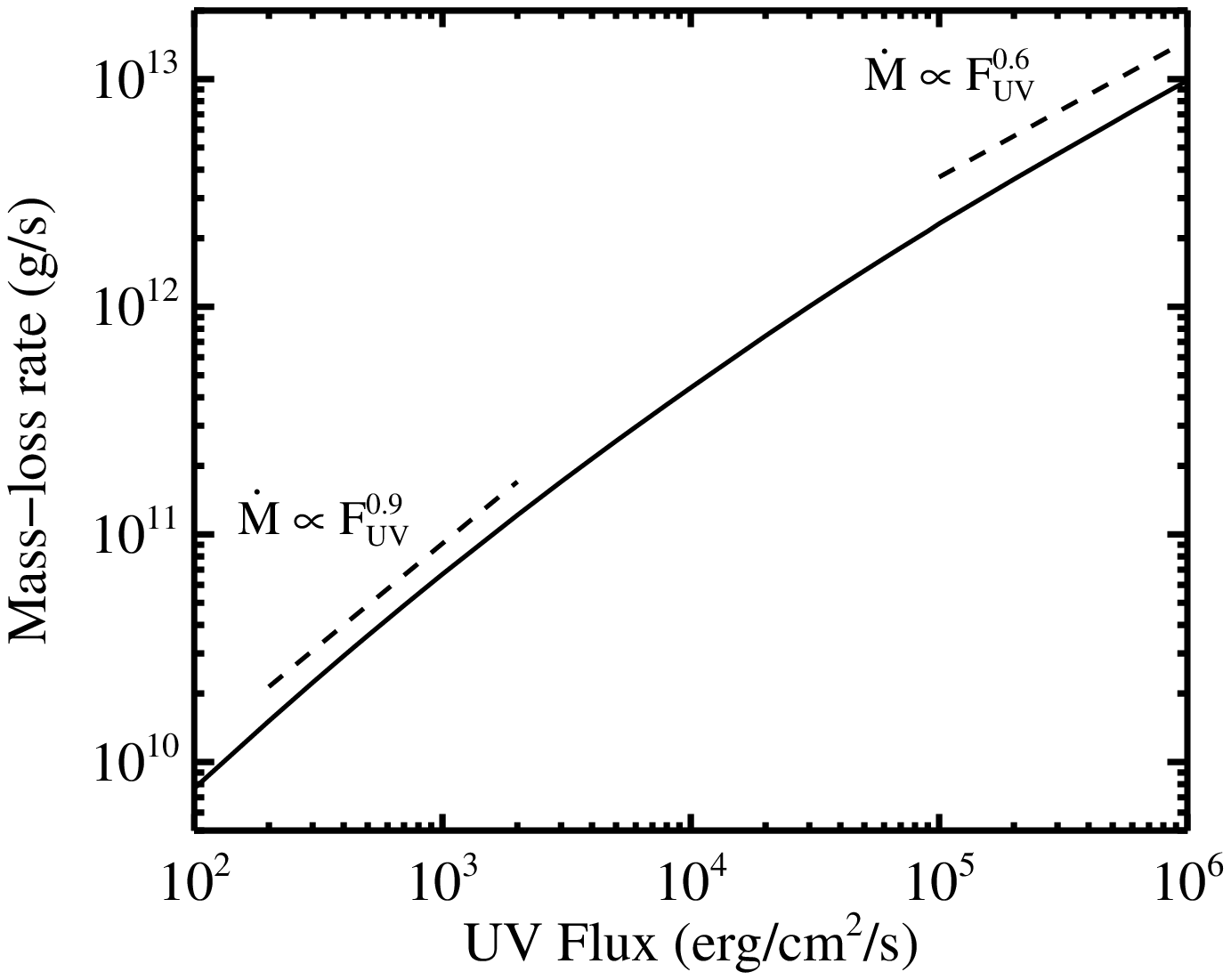}} }
\caption{Maximum mass loss rates as a function of UV flux for a
  $0.7M_{\rm J}$ planet with radius $1.4R_{\rm J}$ located $0.05$ AU
  from a $1M_\sun$ star.  We calculate the mass loss rate along the
  substellar ray connecting the planet to the star, along which escape
  is most easily achieved, and use this solution over the entire
  surface of the planet to calculate the maximum $\dot M = 4\pi
  r^2\rho v$.  More realistic estimates that average the incident
  UV flux over the planetary surface yield mass loss rates
  that are lower by factors of 3--4 (\S\ref{sec-daynight}).
  For $F_{\rm UV} \lesssim 10^4$ erg/cm$^2$/s, $\dot M
  \propto F_{\rm UV}^{0.9}$.  For $F_{\rm UV} \gtrsim 10^4$
  erg/cm$^2$/s, the mass-loss rate increases more slowly as $\dot M
  \propto F_{\rm UV}^{0.6}$. These two regimes correspond to
  ``energy-limited'' and ``radiation/recombination-limited'' flows, as
  explained in \S\ref{sec-twoexplain}.}
\label{fig-mdots}
\end{figure}

We calculate the maximum (i.e., spherically symmetric)
mass loss rate as a function of incident UV
flux and display the result in Figure \ref{fig-mdots}.  For $F_{\rm
  UV}$ less than $\sim$$10^4 \erg \cm^{-2} \s^{-1}$, $\dot M \propto F_{\rm
  UV}^{0.9}$.  For larger UV fluxes, the mass loss rate increases more
slowly as $\dot M \propto F_{\rm UV}^{0.6}$. We discuss the origin of this
difference in \S\ref{sec-twoexplain}.

\section{DISCUSSION}\label{sec-discuss}

In \S\ref{sec-windvsjeans},
we check the validity of our assumption that the flow
is hydrodynamic. In \S\ref{sec-twoexplain}, we explain
how there exist two modes for the wind, and derive analytic expressions for
the mass loss rate that reproduce fairly well our numerical
results. In \S\ref{sec-daynight},
the assumption of spherical symmetry is relaxed;
we estimate the factors by which mass loss rates are reduced
because of variable UV irradiance across the surface
of the planet, and because of variable tidal gravity.
In \S\ref{sec-breeze}, we consider
how the planetary outflow interacts with the ambient stellar
wind---and how such interaction can strongly suppress dayside winds
while energizing nightside winds.
Finally, in \S\ref{sec-radpress},
we assess whether stellar radiation pressure can drive significant
planetary outflows.

\subsection{Hydrodynamic vs. Jeans Escape}\label{sec-windvsjeans}
\begin{figure}
\centering
\scalebox{1.3}{\plotone{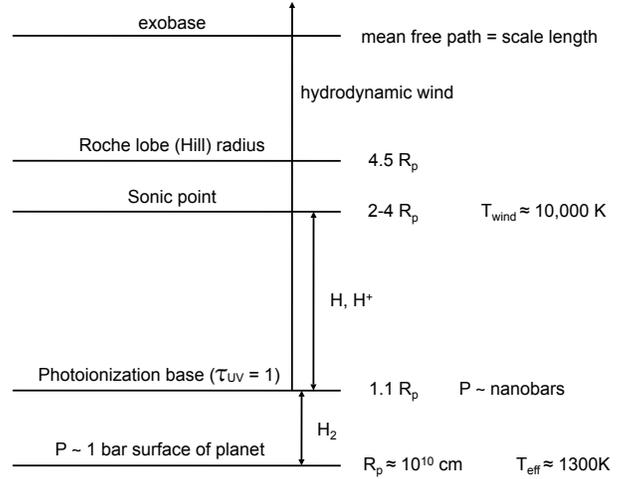}} 
\caption{Summary of important scales in our wind solutions.  For a
  planet with properties inspired by HD 209458b, the $\tau = 1$
  surface to which UV photons penetrate is located where the local pressure
  is measured in nanobars, at about 1.1 times the radius of the 1 bar surface.
  Below this photoionization base, the
  temperature is $\sim$$10^3$ K, close to the effective temperature,
  as befits an atmosphere in radiative equilibrium with the incident stellar
  optical radiation balancing the planet's thermal emission.
  Molecular chemistry (which we do
  not model) is important in this regime.  Our model is valid above
  the $\tau = 1$ surface, where the temperature is
  thermostated to $\sim$$10^4$ K by Lyman-$\alpha$ cooling.
  We demonstrate in Appendix \ref{sec-bcsense} that
  our standard wind solution is insensitive to
  conditions below the $\tau = 1$ surface, justifying our simple model
  which treats only atomic and ionized hydrogen.  The
  sonic point of our wind solutions lies at 2--4 planetary radii,
  with smaller sonic point distances corresponding to higher UV fluxes.
  The Roche lobe radius of the planet at $\sim$$4.5 R_{\rm p}$ is close enough to the
  sonic point radius that tidal gravity is significant.
  Below the exobase, where the mean free path to collisions
  equals the scale height of the atmosphere, the gas behaves as a
  fluid.  Because the exobase is well outside the sonic point of our
  winds, mass loss from hot Jupiters takes the form of a hydrodynamic
  wind rather than Jeans escape.}
\label{fig-scales}
\end{figure}

We have modeled the mass outflow from a hot Jupiter as a hydrodynamic
wind rather than as local Jeans escape.  For this fluid
description to be accurate, the gas must remain collisional until it
is above the sonic point of the flow \citep[p. 377]{ch87}.  In other
words, the exobase---the height at which the scale length $H\equiv
\rho(d\rho/dr)^{-1}$ equals the mean free path to collisions
$\lambda_{\rm mfp} = 1/(n\sigma_{\rm pp})$, where $\sigma_{\rm pp}
\sim 10^{-13}(T/10^4 \K)^{-2}\cm^{-2}$ is the Coulomb cross section for
protons scattering off protons \citep[e.g.,][]{s78}---must lie above the sonic point. This
requirement is easily satisfied by our models.  For our standard model
with $F_{\rm UV} = 450 \erg \cm^{-2} \s^{-1}$, $H/\lambda_{\rm mfp}
\sim 10^4$ at the sonic point.  For $F_{\rm UV} = 5\times10^5 \erg
\cm^{-2} \s^{-1}$, $H/\lambda_{\rm mfp} \sim 10^5$ at the sonic point.
Therefore the gas behaves as a collisional fluid, 
and mass loss rates based on Jeans escape criteria
\citep{lvm+04,y04} are not appropriate.
We agree with \citet{ttp+05} and \citet{g07} that mass loss from hot
Jupiters is hydrodynamic, not ballistic.  Figure \ref{fig-scales}
summarizes the scales in our wind solutions.

Note that for our standard model of the substellar streamline,
the sonic point lies inside the
Roche lobe, and therefore our account of tidal gravity in (\ref{eqn-mom})
is adequate. There may be other streamlines where the sonic point is
outside the Roche lobe. \citet{jel+05} suggest that for such streamlines,
mass loss occurs via ``geometrical blow-off'' and the Jeans escape
criteria of \citet{lvm+04} should be applied.
We disagree; even if the sonic point lies outside the Roche lobe,
gas pressure gradients inside the Roche lobe will drive an outflow, and
wherever the gas remains collisional,
the flow must be solved using the equations of hydrodynamics.

\subsection{Energy-Limited vs. Radiation/Recombination-Limited}\label{sec-twoexplain}
The two regimes for mass loss that our numerical solution
uncovered---$\dot{M} \propto F_{\rm UV}^{0.9}$ at low $F_{\rm UV}$
and $\dot{M} \propto F_{\rm UV}^{0.6}$ at high $F_{\rm UV}$
(Figure \ref{fig-mdots})---can be understood simply.

At low $F_{\rm UV}$, the flow
is largely ``energy-limited.'' Most of the energy deposited 
by photoionizations
as heat---i.e., $\sim$$\varepsilon \pi F_{\rm UV} R_{\rm p}^2$---goes into $PdV$ work,
with little loss to radiation and internal energy changes
(the relative contributions to the energy budget
are given in \S\ref{sec-standardres}).
The $PdV$
work lifts material
out of the planet's gravitational potential well: measured
per unit mass, the work done is
$$
\frac{P \Delta V}{\rho R_{\rm p}^2 H} \sim \frac{P R_{\rm p}^3}{\rho R_{\rm p}^2 H} \sim \frac{\rho g H R_{\rm p}^3}{\rho R_{\rm p}^2 H} \sim \frac{GM_{\rm p}}{R_{\rm p}} \,.
$$
Then the energy-limited mass loss rate is given by
\begin{eqnarray}
\dot{M}_{\rm e-lim} &\sim& \frac{\varepsilon \pi F_{\rm UV}R_{\rm p}^2}{GM_{\rm p}/R_{\rm p}} \nonumber \\
&\sim& 6 \times 10^9 \left( \frac{\varepsilon}{0.3} \right)  \left( \frac{R_{\rm p}}{10^{10}\cm} \right)^3 \left( \frac{0.7 M_{\rm J}}{M_{\rm p}} \right) \nonumber \\
&& \left( \frac{F_{\rm UV}}{450 \erg \cm^{-2} \s^{-1}} \right)  \gm \s^{-1} \,,
\label{eqn-energylimited}
\end{eqnarray}
close to the result found numerically at low $F_{\rm UV}$ (the factor of 5
difference in normalization between Equation \ref{eqn-energylimited} and
the curve shown in Figure \ref{fig-mdots} arises mostly because the latter
takes the substellar UV flux and applies it over all $4\pi$ steradians,
whereas the former averages the UV flux over the surface of the planet---hence
the factor of $\pi$ in equation \ref{eqn-energylimited}).
Energy-limited outflows
were also found by \citet{wdw81} who studied
mass loss in the highly conductive atmospheres of the terrestrial planets.\footnote{Note, however, that \citet{wdw81} reserve the phrase ``energy-limited'' for
use in another context. Nevertheless their Equation (2) is essentially the
same as our Equation (\ref{eqn-energylimited}), the ``energy-limited'' mass loss
rate in the sense that we use the phrase.}

At high $F_{\rm UV}$, the flow is ``radiation/recombination-limited.''
As quantified in \S\ref{sec-highflux},
the input UV power is largely lost to cooling radiation. Radiative losses
thermostat the gas temperature to $T \sim 10^4 \K$.
Under the approximation that the wind is isothermal,
$\dot{M} \sim 4 \pi \rho_{\rm s} c_{\rm s} r_{\rm s}^2$ at the sonic point,
where $c_{\rm s} = [kT/(m_{\rm H}/2)]^{1/2}$ is the isothermal sound speed
(the factor of 2 accounts for the fact that the hydrogen
is nearly completely ionized),
$r_{\rm s} = GM_{\rm p}/(2c_{\rm s}^2)$, and $\rho_{\rm s}$ is the sonic point
density.\footnote{See, e.g., 
\citealt{lc99}.}
Between the
$\tau = 1$ surface and the sonic point,
the density structure is nearly hydrostatic, so that
$$
\rho_{\rm s} \sim \rho_{\rm base} \exp \left[ \frac{GM_{\rm p}}{R_{\rm p}c_{\rm s}^2} \left( \frac{R_{\rm p}}{r_{\rm s}} -1 \right) \right] \,.
$$
In the high flux case, the density $\rho_{\rm base}$ at the $\tau = 1$
photoionization base is $n_{+,{\rm base}} m_{\rm H}$.
That density is determined by ionization equilibrium, which
involves a balance between photoionizations
and radiative recombinations (Figure \ref{fig-5e5terms}):
$$
\frac{F_{\rm UV}}{h\nu_0} \sigma_{\nu_0} n_{0,{\rm base}} \sim n_{+,{\rm base}}^2 \alpha_{\rm rec} \,,
$$
neglecting the order unity flux attenuation at the $\tau = 1$ surface.
The base neutral density
$n_{0,{\rm base}} \sim 1/(\sigma_{\nu_0}H_{\rm base}) \sim m_{\rm H}g/(2\sigma_{\nu_0}kT)$.
This neutral density is fairly insensitive to $F_{\rm UV}$ (see beginning of
\S\ref{sec-model} and compare Figures \ref{fig-450nfptau}
and \ref{fig-5e5nfptau}),
and so we conclude that the radiation/recombination-limited mass loss rate is
given by
\begin{equation}
\dot{M}_{\rm rr-lim} \sim 4 \times 10^{12} \left( \frac{F_{\rm UV}}{5 \times 10^5 \erg \cm^{-2}\s^{-1}} \right)^{1/2} \gm \s^{-1}\,,
\label{eqn-rrlimited}
\end{equation}
similar to the answer found numerically at high $F_{\rm UV}$.

\subsection{Spherical Asymmetry: Day/Night and Tidal Gravity}\label{sec-daynight}
The mass loss rates given in all our plots
are upper limits because they take our 1D solution
for the substellar streamline and apply it over $4\pi$ steradians.
The mass flux is maximized for the substellar streamline because
the substellar point receives the maximal UV flux, and because the tidal
gravity term weakens the planet's gravity most
along the line joining the planet to
the star. We now discuss the extent to which the actual
mass loss rate is reduced because of day/night differences in the
received UV flux, and because of the directional dependence of
tidal gravity. For simplicity we discuss these effects separately,
as if they could be isolated from one another.

Ignoring for the moment differences between day/night external
boundary conditions---these are actually significant and dealt with in
\S\ref{sec-breeze}---day/night differences would be erased in the
extreme case that horizontal winds redistribute photoionized plasma
from the dayside of the planet to the nightside, on a timescale
shorter than the wind's radial advection time of a few hours.  In this
case, the ``nightside wind'' would blow just as strongly as the
``dayside wind.''  In reality, if the dayside wind blows freely (see
\S\ref{sec-breeze} for important reasons to believe that it may not),
the timescale for horizontal advection is at least as long as that for
radial advection, since the distances travelled in both cases are
several $R_{\rm p}$, the radial speed is supersonic, and the
horizontal speed is at most sonic. So if the dayside wind blows, the
nightside wind is likely muted.  The inverse is also true---see
\S\ref{sec-breeze}.

In the opposite limit of no horizontal redistribution, the mass loss
rates we have calculated would be reduced by a factor of
$(1/2)\int_0^{\pi/2}\sin \theta (\cos \theta)^{\gamma}d\theta =
1/[2(\gamma+1)]$, where $\theta$ is the angle measured from the
substellar ray ($\theta = 0$ points along the substellar ray, while
$\theta = \pi /2 $ defines the terminator dividing day from night),
and the $(\cos \theta)^{\gamma}$ factor accounts for how the planetary
mass flux scales with incident stellar flux (which itself scales as
$\cos\theta$).  For energy-limited flows, $\gamma = 0.9$, and for
radiation/recombination-limited flows, $\gamma=0.6$ (see Figure
\ref{fig-mdots}).  The reduction factor equals 1/3.8 = 0.26 and 1/3.2
= 0.31 in the two respective cases. If there is horizontal
redistribution, we do not expect the reduction factors to change
appreciably from these values. To first order, redistribution of
plasma simply redistributes the wind over the planetary
surface. Changes in the total mass loss rate are expected to be of
second order. For example, if the wind is strictly energy-limited
(\S\ref{sec-twoexplain}), the mass loss rate depends only on the
amount of UV radiation intercepted by the planet, and is independent
of the degree of redistribution.

To get a sense of how much the mass loss rate is reduced because tidal
gravity does not point parallel to all streamlines emanating from the
planet, we eliminate the tidal term from the force equation
(\ref{eqn-mom}) and re-solve the fluid equations.  We obtain a mass
loss rate (multiplying by $4\pi$ steradians) that is 0.79 $\times$ our
standard model result.  This can be considered a maximum
reduction factor for the directional dependence of 
tidal gravity insofar as its effect can be isolated.

\begin{figure}
\centering
\vspace{-0.1in} 
\hbox{\hspace{-0.2in}\scalebox{1.25}{\plotone{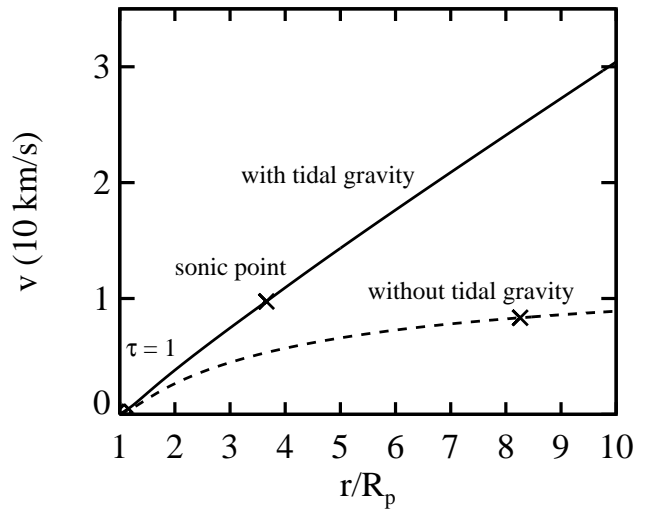}}}
\caption{Wind velocity as a function of altitude
  for our standard model with tidal gravity (solid
  curve) and without (dashed curve).
  Tidal gravity alters the velocity profile of the wind, allowing the flow
  to attain higher speeds.  It also moves the wind's sonic point to lower
  altitude.}
\label{fig-450v_tnt}
\end{figure}

Though tidal gravity does not change the calculated mass loss rate
appreciably, it is nonetheless important for two reasons.  First,
tidal gravity alters the velocity structure of the wind,
allowing it to accelerate to larger speeds (Figure \ref{fig-450v_tnt}).
Second, tidal gravity moves the
location of the sonic point to lower altitude, allowing
it to remain within the Roche lobe along the substellar ray and
helping validate our 1D treatment.

To summarize this subsection, if the dayside transonic wind blows freely,
our crude accounting for the
directional dependences of UV irradiance and of tidal gravity suggest
that actual mass loss rates are lower than our plotted
values by factors of $\sim$4. We now turn to the question
of whether the dayside wind can actually blow, when faced
with the considerable pressure of the host star wind.

\subsection{Colliding Winds and Breezes: Trading the Dayside Wind for the Nightside Wind}\label{sec-breeze}

\begin{figure}
\centering
\vspace{-0.8in}  
\hbox{\hspace{-0.45in}\scalebox{1.6}{\plotone{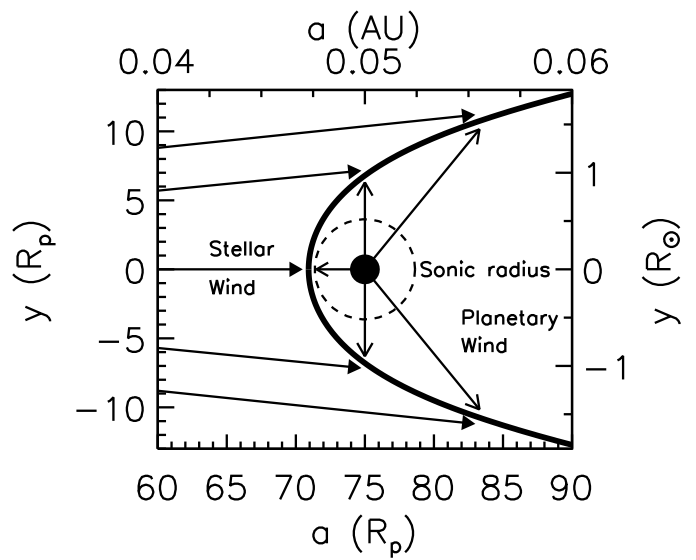}} }
\vspace{-0.5in}  
\caption{Cartoon of the standing bow shock (heavy curve)
where planetary and stellar winds collide and mix. For this figure only,
in order to compute the location of the bow shock,
we assume for simplicity that the winds are spherically symmetric, have zero
thermal and magnetic pressure, and have constant velocity (we neglect
Coriolis, centrifugal, and gravitational forces---we have not directly
used one of our wind models). For the stellar wind, we
assume a mass loss rate of $5 \times 10^{11} \gm \s^{-1}$ and
a velocity of 400 km/s. For the planetary wind, we take
$3\times 10^{10} \gm \s^{-1}$ and 20 km/s, values informed by
our wind models. The bow shock is given
by the condition that the bulk momentum fluxes (``ram'' pressures)
normal to the shock
balance \citep[e.g.,][]{lmm90}. Such a shock does not influence
the planetary wind upstream if it is located outside the sonic radius,
as is drawn. In reality, this condition may not be satisfied because
of the overwhelming total pressure of the stellar wind.
}
\label{fig-cartoon}
\end{figure}

The planetary wind does not exist in vacuum. The dayside wind 
blows into the plasma streaming from the star: the stellar wind.
Stellar and planetary winds collide and mix in a standing bow
shock surrounding the planet.
The situation is analogous to the colliding winds of massive stellar
binaries \citep{lmm90,sbp92}.
Figure \ref{fig-cartoon} supplies a cartoon illustration.
However, as stressed in the figure caption and in the discussion below,
numerous assumptions underlie this cartoon, and reality is likely
to look substantially more complicated.

At the stand-off shock, the total pressures of the two flows
balance. 
The total pressures include contributions from ``ram'' ($\rho v^2$),
thermal, and magnetic pressure, but should include only components
that are normal to the shock front (the shocks are oblique).
The total pressure of the planetary wind decreases monotonically with distance
from the planet.
If the total stellar wind pressure $P_{\ast,{\rm tot}}$ near the planet is
less than the total planetary wind pressure $P_{\rm tot}(r_{\rm s})$
at the planetary wind's sonic
point, then the stand-off shock will be located downstream of the planet's
sonic point. There it cannot influence the planetary
flow upstream, for the same reason that one cannot shout upstream
in a supersonic flow and be heard.\footnote{Though
magnetic disturbances can propagate upstream if the stand-off shock
occurs inside the planetary wind's Alfv\'enic and fast magnetosonic points
(see, e.g., \citealt{wd67} for a theory of magnetized winds).}
In this case the transonic wind
solution that we have obtained for $r < r_{\rm s}$---and in particular
our computed mass flux---would remain unchanged.

\begin{figure}
\centering
\vspace{-0.1in}
\scalebox{1.2}{\plotone{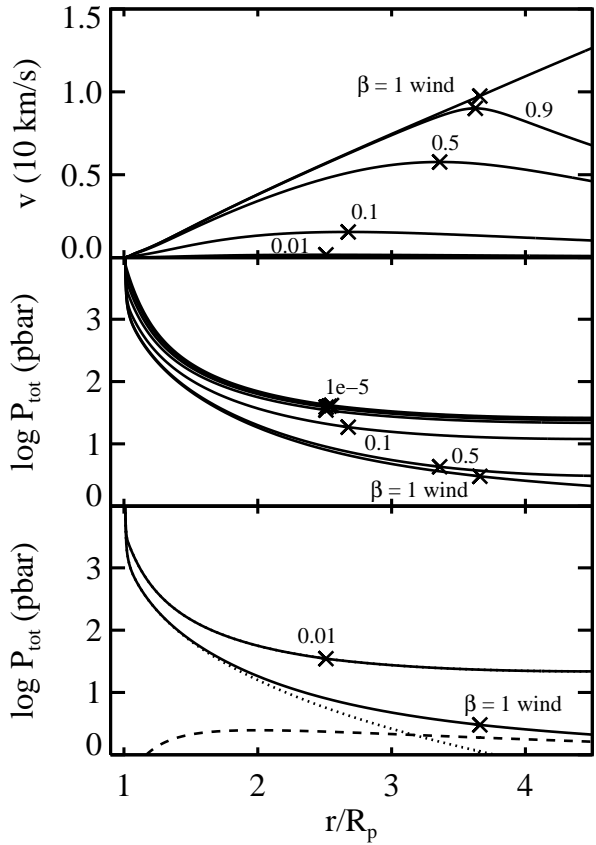}} 
\caption{
Velocity and total pressure ($P_{\rm tot} = \rho v^2 + P$) profiles
for our unique transonic wind ($\beta = 1$) and a family of breezes ($\beta<1$
as labelled), for standard model parameters. The x symbol marks $r=r_{\rm s}$,
which for the wind represents the sonic point, and for breezes
represents the point of maximum velocity. At infinity, breezes
remain pressurized while the wind does not. In the middle panel,
profiles not labelled include $\beta = 10^{-2}$, $10^{-3}$, and $10^{-4}$.
Note how these profiles hardly change as $\beta$ decreases---the solution
is approaching that of a $v=0$ hydrostatic atmosphere.
The bottom panel shows that for the wind, thermal pressure
dominates at depth (dotted line), while $\rho v^2$ ``ram'' pressure 
dominates at altitude (dashed line). For the $\beta = 0.01$ breeze shown,
thermal pressure dominates everywhere.
For the most part, $r_{\rm s}$ shrinks as ever slower breezes (smaller $\beta$) 
are considered. The exception occurs for
$\beta \lesssim 10^{-2}$ (middle panel), and is an artifact of our
choice to compute $\tau(r_{\rm s})$ out to the fixed Roche lobe radius $R_{\rm Roche}$;
$\tau(r_{\rm s})$ increases with decreasing $\beta \lesssim 10^{-2}$.
}
\label{fig-breeze}
\end{figure}

What are $P_{\ast,{\rm tot}}$ and $P_{\rm tot}(r_{\rm s})$ for a hot Jupiter
orbiting a main-sequence, solar type star? For the former quantity
we are guided by observations
of the solar wind. At $a = 0.05 \AU = 10 R_{\odot}$,
the solar wind has accelerated to nearly its maximum speed \citep{mab97,mam03}.
Approximating the stellar wind velocity as constant from 0.05 to 1 AU,
we scale characteristic spacecraft measurements (SOHO\footnote{http://umtof.umd.edu/pm/}; Ulysses\footnote{http://swoops.lanl.gov/recentvu.html})
of the solar wind density and velocity at 1 AU
to estimate the corresponding density $n_{\ast}$ and velocity
$v_{\ast}$ at 0.05 AU:
$n_{\ast} \sim 6 (1/0.05)^2 \, {\rm protons} \cm^{-3}$ and
$v_{\ast} \sim 400 \km \s^{-1}$.
We take the local proton temperature 
to be $T_{\ast} \sim 10^6\K$ (the electron temperature is several
times lower; \citealt{mab97}), and the
magnetic field strength to be $B_{\ast} \sim 0.01$ G at 10 stellar radii
\citep{bam98}.
Then $P_{\ast,{\rm tot}} \sim n_{\ast} m_{\rm H} v_{\ast}^2 + n_{\ast}kT_{\ast}
+ B_{\ast}^2/(8\pi) \sim 10$ picobars,
with the ram and magnetic pressures dominating.
This is an overestimate insofar as we are not taking the components
of the ram and magnetic pressures that are relevant perpendicular to
the shock front.
Compare this to $P_{\rm tot}(r_{\rm s}) \sim 3$ pbars, as calculated for our
standard model of the substellar streamline
(see Figure \ref{fig-breeze}), neglecting
the unknown magnetization of the planet.\footnote{
It has been speculated
that the magnetic fields of hot Jupiters are weaker than
that of Jupiter, since the rotation of a hot Jupiter
is tidally slowed to nearly the orbital
period of three days (exact synchronization is not possible if the
planet lacks a permanent quadrupole moment; 
see, e.g., the textbook by \citealt{md00}),
whereas Jupiter's rotation period is ten hours. 
\citet{b47} speculates that planetary magnetic moments scale linearly with
rotation frequency; if so, the surface field on a hot Jupiter at $r = R_{\rm p}$
would be $\sim$1 G. If the field falls as a dipole to
$r = r_{\rm s} \approx 3 R_{\rm p}$, it would add of order 10 pbar
to $P_{\rm tot}(r_{\rm s})$.}
Given the uncertainties,
the most we can say is that $P_{\ast,{\rm tot}}$ and $P_{\rm tot}(r_{\rm s})$
are comparable.
While a transonic wind solution along certain (not necessarily substellar)
streamlines may yet be possible, it is also possible---indeed
even likely, given the fact that the solar wind can gust to
values of $n_{\ast}$ and $v_{\ast}$ several times higher than the ones
we have used---that the stellar wind squashes the planetary
dayside outflow down to a subsonic breeze, or even stops dayside
photoevaporative mass loss completely.

A breeze would still blow a bubble in the stellar wind, like
that drawn in Figure \ref{fig-cartoon}, only smaller. But because
a breeze is subsonic, it would not traverse a shock at the bubble boundary.
Unlike the case for
the supersonic wind, a breeze is in causal contact with the flow
at the boundary; i.e., conditions at the boundary influence, via sound waves,
conditions in the interior.
The radial velocity would decrease smoothly to zero at the bubble boundary,
giving rise to large pressure gradients that would drive
flows parallel to the boundary (by analogy to subsonic flow around
a blunt obstacle such as a hard sphere). The flow inside the bubble would
be nonradial and multidimensional.
Unfortunately, we cannot easily capture such behavior
with our 1D model; if we tried to model a breeze by imposing
a boundary condition of zero radial velocity at finite
distance from the planet, while insisting simultaneously
on a non-zero mass loss rate, our simple continuity relation
(\ref{eqn-con}) would yield infinite density.

\begin{figure*}
\centering
\scalebox{1.1}{\plotone{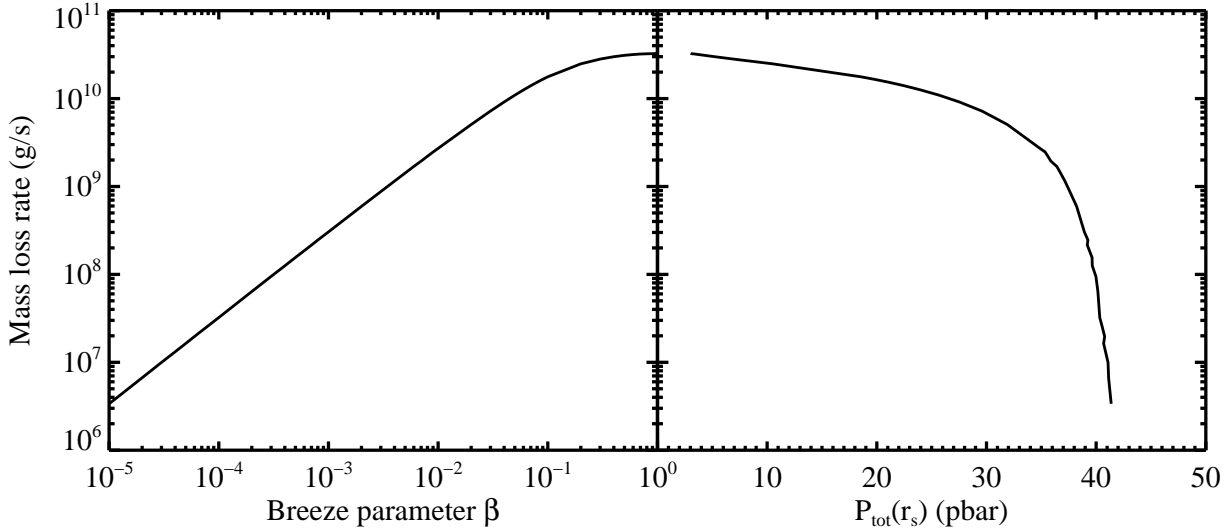}} 
\caption{Mass loss rates (assuming $4\pi$ symmetry) for our
breeze solutions. Mass loss rates become arbitrarily small
as the breeze parameter $\beta$---which sets the maximum breeze
velocity at $r_{\rm s}$---decreases. Interestingly, the
total pressure $P_{\rm tot}(r_{\rm s})$ asymptotes to
a maximum value of $\sim$40 pbar as ever slower breezes are considered
(see also Figure \ref{fig-breeze}). 
The existence of a maximum pressure
suggests that if the stellar wind exerts an external pressure
in excess of $\sim$40
pbar (for our standard parameters and boundary conditions),
it will shut down the dayside planetary outflow completely.
}
\label{fig-breezemdot}
\end{figure*}

Nevertheless, we can get a sense
of how much the dayside mass loss rate might be reduced in the presence
of external pressure,
by re-running our relaxation code with boundary condition (\ref{eqn-bc1})
replaced by
$[v^2 = \beta \gamma kT/\mu]_{r_{\rm s}}$, where $\beta < 1$ is our breeze
parameter.
Note that $r_{\rm s}$ in this case is no longer the sonic point but is
instead the location where the flow stops accelerating and starts decelerating.
Resulting breeze solutions are shown in Figure \ref{fig-breeze},
with the corresponding mass loss rates plotted in Figure \ref{fig-breezemdot}.
While the total pressure $P_{\rm tot}$ for the $\beta=1$ wind solution
decreases to zero at large distance, $P_{\rm tot}$ for a $\beta < 1$ breeze
asymptotes to a finite value. As $\beta$ decreases,
the breezes blow more slowly, approaching the $v=0$
limiting solution for a hydrostatic atmosphere.
Once $\beta \lesssim 10^{-2}$ ($v \lesssim 0.1 \sqrt{\gamma kT/\mu}$),
the $P_{\rm tot}$ profiles hardly change.
We take as a reference pressure $P_{{\rm tot}}(r_{\rm s})$;
the location $r_{\rm s}$ is a sensible one to examine since for the most part
it decreases as $\beta$ decreases, mimicking the shrinking of the planetary
bubble with stronger stellar winds.
Over the entire family of breezes, the reference pressure $P_{\rm tot}(r_{\rm s})$
achieves a maximum value of $\sim$40 pbar for our standard parameters
and boundary conditions. (Recall that an isothermal hydrostatic
atmosphere has a finite pressure at infinity.)

To the extent that our 1D solution lends insight into the 3D breeze
problem, we conclude the following: if the stellar wind pressure
$P_{\ast,{\rm tot}} \lesssim 3$ pbars, the planet emits a full-fledged
transonic wind; if $P_{\ast,{\rm tot}}$ lies between $\sim$3 and
$\sim$40 pbars, then the planet emits a dayside breeze with a finite
mass loss rate; but if $P_{\ast,{\rm tot}} \gtrsim 40$ pbars, then the
planet's dayside atmosphere is forced to be radially hydrostatic, with
the stellar wind penetrating to the depth where pressure balance is
achieved.  Figure \ref{fig-breezemdot} informs us that tiny increases
in $P_{\rm tot}(r_{\rm s})$ above 35 pbars correspond to enormous
reductions in $\dot{M}$.  At the same, if $P_{\rm tot}(r_{\rm s})$ is
below 35 pbar, the mass loss rate is $\sim$$10^{10} \gm \s^{-1}$, to
within a factor of about 3. Thus, the stellar wind can act essentially
as an on-off switch---at times permitting the planet to lose mass from
its dayside at near maximum rates of order $10^{10} \gm \s^{-1}$, and
at others shutting down dayside mass loss completely.  In the case of
hot Jupiters orbiting main-sequence stars, given how the stellar
outflow pressure is of order 10 pbar and how it may increase
dramatically during violent coronal mass ejections, we expect this switch to
flip back and forth with stellar activity.  We cannot rule out the
possibility that the dayside switch might even be off the majority of
the time.

Even when the dayside switch is off, however, we do not expect
mass loss to shut down completely. As the dayside outflow weakens,
the nightside wind should strengthen in proportion. That is because
the UV energy that the planet absorbs must be spent one way or another;
if it is prevented from doing $PdV$ work against gravity on the dayside,
it will instead power horizontal winds which carry heat to the nightside.\footnote{\citet{kam07,kas+07} construct a thermospheric circulation model for hot Jupiters similar to HD 209458b but located further than 0.16 AU from their host stars.  They argue that at these distances, atmospheric temperatures remain low enough that molecular hydrogen never dissociates, H$_3^+$ dominates cooling, and the atmosphere remains in vertical hydrostatic equilibrium.  Similar circulation models could be made for the thermospheres of atomic hydrogen of closer-in planets, in the case where the dayside atmosphere is forced to be hydrostatic by the stellar wind.}
Shielded from the stellar wind, the nightside is free to emit
its own wind. Thus, the total mass loss rate might remain roughly constant
at $\sim$$10^{10}\gm \s^{-1}$, even when the stellar wind gusts (assuming
that no other energy sink becomes active).

In concluding that nightside winds will blow when dayside mass loss has been quenched, we have implicitly assumed that the planet resides outside of the region where the dipolar component of the star's magnetic field dominates.  Inside this region---i.e., inside the star's Alfv\'en radius---most magnetic field lines are not blown open by the stellar wind.  Closed, poloidal field lines can encage both dayside and nightside winds, confining atmospheric escape to magnetic flux tubes in the vicinity of the planet's poles.  The detailed structure of the solar magnetosphere has not been conclusively established.  
Some models \citep[e.g.,][]{mab97,mam03} place
the solar Alfv\'en radius near 10 $R_{\odot}$, near the orbit of a hot
Jupiter. However, the three-dimensional model
of \citet[][their ``DQCS'' model with Q = 1.5]{bam98} states
that at $10 R_{\odot}$,
field lines are primarily radial, and by implication,
the Alfv\'en surface lies at smaller radius.

\begin{figure}
\centering
\hbox{\hspace{-0.2in}\scalebox{1.25}{\plotone{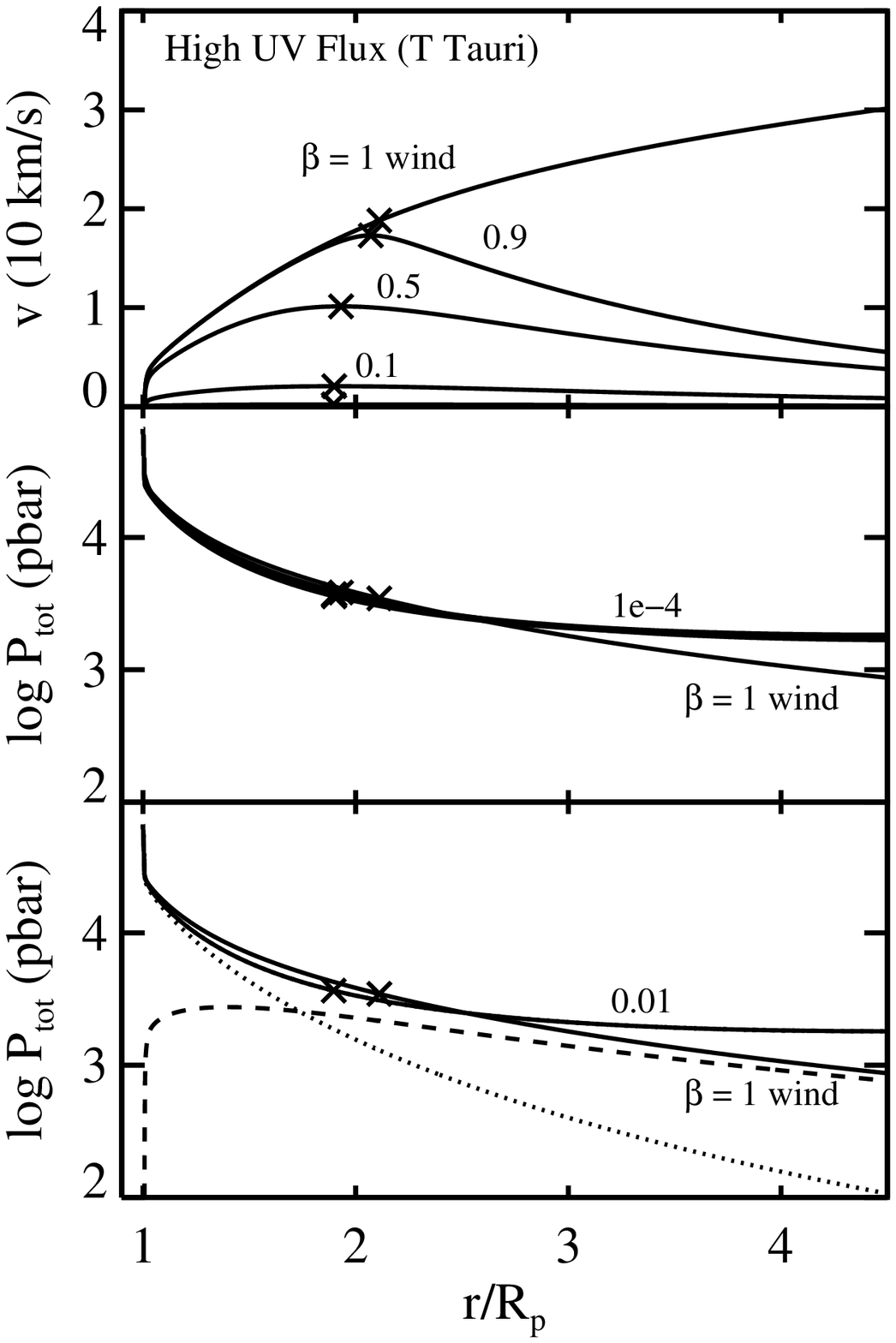}} }
\caption{
Same as Figure \ref{fig-breeze}, but for the high
flux case where $F_{\rm UV} = 5 \times 10^5 \erg \cm^{-2} \s^{-1}$.
Over all wind and breeze solutions,
the reference pressure $P_{\rm tot}(r_{\rm s})$ does not exceed
$\sim$4000 pbar (middle panel). Since T Tauri stellar wind pressures
are about three orders of magnitude greater than this (see text), 
we expect the daysides of hot Jupiters orbiting pre-main-sequence stars
to emit neither winds nor breezes, but to have radially hydrostatic
atmospheres. Their nightsides, however, may well be losing mass
at a rate of order $10^{12} \gm \s^{-1}$, since we expect horizontal
winds to carry hot photoionized plasma from the dayside to the nightside.
}
\label{fig-breezes_5e5}
\end{figure}

What about for hot Jupiters orbiting T Tauri stars? To what extent are
their dayside
outflows squashed by the highly pressurized winds emitted by young active
stars? If magnetospheric truncation of T Tauri accretion disks sets the final orbits of inwardly migrating hot Jupiters \citep{lbr96}, these planets likely also
reside near the Alfv\'en radii of their host star magnetospheres, where the stellar ram and magnetic pressures are comparable.  This conclusion is compatible with observations of mass loss rates from T Tauri stars in the range $10^{-9}$ to $10^{-7}M_\sun/$yr \citep{ecs+87} and T Tauri surface fields that are $10^3$ times stronger than their main-sequence counterparts \citep[e.g.,][]{jvk99}.
We therefore estimate that 
$P_{\ast,{\rm tot}} \sim 10 \times (10^3)^2$ pbar $\sim 10$ $\mu$bar.
Let us compare this pressure to the pressures characterizing planetary outflows.
Figure \ref{fig-breezes_5e5} shows a family of breeze solutions
for a T Tauri-like UV flux of $F_{\rm UV} = 5 \times 10^5 \erg \cm^{-2} \s^{-1}$.
It shows that $\max(P_{{\rm tot}}(r_{\rm s})) \approx 4$ nanobar
$\ll P_{\ast,{\rm tot}}$.
Thus, it seems safe to conclude that 
T Tauri stellar winds completely stifle dayside winds from hot
Jupiters.

It is not clear whether hot Jupiters around T Tauri stars reside inside or outside their host stars' Alfv\'en radii.  If stellar rotation rates are locked to disk rotation rates at Alfv\'en radii (``disk locking''; e.g., \citealt{hm05a}), this question reduces to whether the planet orbits inside or outside the corotation circle.  The current measured rotation period of HD 209458 of $\sim$12.4 days
\citep[][and references therein]{wnh+05} yields a corotation radius of $\sim$0.1 AU.  However, T Tauri stars typically rotate faster than their main-sequence counterparts \citep[e.g.,][]{hhs+86,jg02}.  Furthermore, observations of T Tauri magnetic fields do not consistently confirm models in which the closed field extends to the corotation radius \citep[and references therein]{jgd08}.  

Given these uncertainties, we acknowledge two possibilities.  Winds from giant planets located well inside the magnetospheres of their T Tauri host stars will be confined on both day and night sides by magnetic pressure, blowing only along polar flux tubes.  Hot Jupiters located in open field line regions experience quenching only on the day side---they lose mass strictly by night.

\subsection{Radiation Pressure}\label{sec-radpress}
Can stellar radiation pressure drive substantial planetary mass loss?
Neutral H atoms absorbing stellar
Ly$\alpha$ photons feel a radiation pressure force $F_{{\rm
  Ly}\alpha}\sigma_{{\rm Ly}\alpha}/c$, where $c$ is the speed of light.
We use the solar Ly$\alpha$ flux scaled to $a = 0.05
\AU$, $F_{{\rm Ly}\alpha} \sim 2.4\times 10^3$erg/cm$^2$/s \citep{wtr+00}.
If the line were only
thermally broadened with a velocity of $\sim$$10\km \s^{-1}$, the
absorption cross section at line center would be $\sigma_{{\rm
Ly}\alpha} \sim 6 \times 10^{-14} \cm^2$.  To account for extra
Doppler broadening from a range of bulk velocities extending up to
$v_{\rm orb}$, we lower the thermally broadened cross section by
another factor of 10: $\sigma_{{\rm Ly}\alpha} \sim 6 \times 10^{-15}
\cm^2$. Putting it all together, we find that the radiation pressure
force is comparable to that of stellar gravity,
$GM_{\ast}m_{\rm H}/a^2$, in agreement with the result of \citet{vld+03}.  
But, radiation pressure of this magnitude would require of order an orbital period (days) to accelerate H atoms to velocities of $v_{\rm orb}\sim 100 \km \s^{-1}$.  The observations, in contrast, require this acceleration to occur over a few hours, the time to travel $\sim$$10 R_{\rm p}$, the size scale probed by the transit.  Furthermore, atoms are subject to Ly$\alpha$ radiation pressure for only several hours before they are photoionized.  These considerations imply that a larger Ly$\alpha$ flux than we have assumed would be required to accelerate neutral hydrogen to $\sim$100 km/s \citep{hes+08}.

Nonetheless, we assume optimistically
that radiation pressure can in fact accelerate H atoms to $\sim$$v_{\rm orb}$ and estimate the
resultant mass loss rate. Radiation pressure can only act on gas that
is optically thin to Ly$\alpha$ photons.
The column density to optical depth unity is $N_{{\rm
    Ly}\alpha} = 1/\sigma_{{\rm Ly}\alpha} \sim 2 \times 10^{14}
\cm^{-2}$.  Hydrogen is accelerated off the planet's limb; the area of
the annulus presented by hydrogen within the planet's Roche lobe is
$\sim$$R_{\rm Roche}^2$. This hydrogen is removed every $R_{\rm
  Roche}/v_{\rm orb}$ time. Putting it all together, we estimate a
mass loss rate due to radiation pressure of
\begin{equation}
\dot{M}_{\rm rad. press.} \sim \frac{N_{{\rm Ly}\alpha}R_{\rm Roche}^2m_{\rm H}}{R_{\rm Roche}/v_{\rm orb}} \sim 10^8 \gm \s^{-1} \,,
\end{equation}
two orders of magnitude lower than the mass loss rates we have computed for
photoionization-heated, hydrodynamic winds. 
We conclude
that radiation pressure is not a significant driver of planetary outflows
when compared against photoionization.

\citet{vld+03}
claim a mass loss rate of $10^{10} \gm \s^{-1}$ based on
line-driven acceleration to large blue-shifted velocities
of gas that has been raised, presumably by other means,
to the altitude of the Roche lobe. Their claim
is based in part on an assumed
hydrogen density of $n_0 \sim 2\times 10^5 \cm^{-3}$
at a distance of $r = R_{\rm Roche}/2$. But the corresponding column
$n_0 R_{\rm Roche} / 2 \sim 4 \times 10^{15} \gg N_{{\rm Ly}\alpha}$ would
be optically thick to Ly$\alpha$ photons ($\tau_{{\rm Ly}\alpha} \sim 20$). 

\section{SUMMARY AND COMPARISON WITH OBSERVATIONS}\label{sec-summary}

We have presented a simple model of atmospheric escape from hot
Jupiters, as driven by photoionization heating by stellar UV
radiation.  To calculate the steady-state structure of the planetary
wind, we employed a relaxation code to solve the equations of
ionization balance and of mass, momentum, and energy conservation.  We
imposed two-point boundary conditions that allowed us to find the
unique wind solution that transitions from subsonic to supersonic
velocities. 
Tidal gravity is important to include in our momentum equation because 
it alters the entire velocity structure of the wind, generating higher
velocities and helping to lower
the wind's sonic point to within the Roche lobe, at
least for the substellar streamline.
Photoionization heating is balanced by a combination of $PdV$ work and
cooling by Lyman-$\alpha$ radiation.  The latter serves as a
thermostat, limiting planetary wind temperatures to $\sim$$10^4$ K
over four decades in incident UV flux.  Notably, conductive transport
of energy is not important, by contrast to 
the thermospheres of terrestrial planets.

Our assumption that mass loss from hot Jupiters takes the form of
hydrodynamic winds rather than local Jeans escape is
self-consistent: in our solution, escaping gas remains collisional at
the sonic point. Showing that this condition is satisfied at the sonic
point is sufficient because the flow inside the sonic point is denser
and hence still more collisional, while the flow outside the sonic
point is supersonic and so cannot influence the flow inside---in a
supersonic flow, one cannot shout upstream and be heard.  
We agree with \citet{ttp+05} and \citet{g07}
that the flow is never in a regime where Jeans escape considerations
are relevant (see \S\ref{sec-windvsjeans}).

We find that a planet similar to the transiting hot Jupiter HD 209458b
loses mass at a maximum rate of $\sim$$10^{10} \gm \s^{-1}$ for a UV
flux characteristic of low to moderate solar activity.
For a UV flux near the Lyman edge that is $\sim$2--3 times larger, as obtains during
solar maximum \citep[][and references therein]{web+05,lwm+03}, the mass loss rate is
$\sim$$2 \times 10^{10}\gm \s^{-1}$ (we have shown that
at these fluxes, mass loss
is energy-limited and scales nearly
linearly with UV flux). This maximum mass loss rate is two orders
of magnitude lower than the $10^{12}\,{\rm g/s}$ quoted by \citet{lsr+03}, and inconsistent with the claim by \citet{bsc+04}
that planets lighter than 1.5 Jupiter masses at 0.05 AU from their
stars evaporate entirely in $< 5$ Gyr.  We agree with \citet{ttp+05}
(who idealize the wind as neutral),
\citet{y06}, and \citet{g07} that the current mass loss rate from HD
209458b causes the planet to lose at most 1\% of its
mass over its 5 Gyr age.  

What explains the factor of 100 discrepancy between our maximum
mass loss rates and the mass loss rates given by
\citet{lsr+03} and \citet{bsc+04}?  These
authors, inspired by \citet{wdw81} who model mass loss from
the highly conductive atmospheres of terrestrial planets,
posit that outflows from hot Jupiters are
energy-limited.  Energy-limited flows are those for which a fixed
fraction of the stellar UV radiation incident upon a planet's surface
goes towards driving gas out of the planet's gravitational well.  As
we have shown, conduction is not important in the winds emitted by hot
Jupiters, and therefore the details of the calculation by
\citet{wdw81} are not transferable.  Nonetheless, we find that for
$F_{\rm UV} \lesssim 10^4 \erg \cm^{-2} \s^{-1}$, as obtains
for hot Jupiters orbiting main-sequence solar analogs,
hot Jupiter winds are indeed nearly energy-limited.
The energy-limited mass loss rate can be written as
\begin{equation}
\dot M_{\rm lim} \approx \frac{\varepsilon F_{\rm UV} \times \pi r_1^2}{G M_{\rm p}/r_0}
\end{equation}
\citep[][see also our Equation \ref{eqn-energylimited}]{wdw81}, where gas is bound to the planet below radius $r_0$ and
the bulk of incoming UV radiation is absorbed at $r_1$.  
The difference between our calculated $\dot M$ and that derived by
\citet{lsr+03} arises from two factors. First, we include a heating
efficiency $\varepsilon < 1$, which must account at least
for the energy lost to
photoionizing atoms. In our simple model, $\varepsilon \approx 0.3$ (recall Equation \ref{eqn-Q0}).
Second, and more significantly, our calculation places $r_0$ and $r_1$
closer to $1.1 R_{\rm p}$ (see beginning of \S\ref{sec-model}), the
location of the $\tau=1$ surface to photoionization.  \citet{lsr+03}
take instead $r_0 \approx r_1 \approx 3 R_{\rm p}$ by applying
detailed formulae derived by \citet{wdw81}.  These formulae are not
appropriately applied to the photoionized upper atmospheres of hot
Jupiters, because in these environments conductive cooling
is not significant. \citet{g07} reaches essentially the same conclusion;
see his section 3.5.

For high UV fluxes $\gtrsim 10^4 \erg \cm^{-2} \s^{-1}$, like those
incident upon hot Jupiters orbiting active T Tauri stars, mass loss
ceases to be energy-limited. Most of a fast photoelectron's energy
is lost to collisionally excited Ly$\alpha$ radiation.  By contrast to the
case at low $F_{\rm UV}$ where photoionizations are balanced by gas
advection and the neutral gas fraction remains of order unity at
altitude, at high $F_{\rm UV}$ photoionizations are balanced by
radiative recombinations.  The transition to a nearly completely
ionized flow is sharp, and the overall wind structure is reminiscent
of a classic expanding \ion{H}{II} region. Whereas in the
energy-limited regime $\dot{M}$ is expected to scale as $F_{\rm
  UV}^1$, in the radiation/recombination-limited regime $\dot{M}$ is
expected to scale as $F_{\rm UV}^{1/2}$, essentially because that is
how the number density of ionized atoms scales in a plane-parallel Str\"omgren
slab. Our detailed numerical model yields power-law indices of 0.9 and
0.6 at low and high UV flux, respectively.

We have demonstrated that above the $\tau = 1$ surface to
photoionization, our wind model is insensitive to our chosen boundary
conditions.  This helps to justify our neglect of hydrogen molecular
chemistry. The uncertainty generated by this omission
is embodied in our choice of the base radius $r_{\rm min}$,
which by definition is that radius where hydrogen is predominantly
atomic. Without modelling the chemistry of H$_2$, we cannot be sure
where $r_{\rm min}$ is located. But we have shown that
we can estimate it to sufficient
accuracy (about 10\%; see the order-of-magnitude discussion
at the beginning of \S\ref{sec-model})
that our mass loss rates are uncertain by at most
factors of a few (see Figure \ref{fig-bcrmin} of Appendix \ref{sec-bcsense}).
The bulk properties of our wind solutions are in good agreement with
those of \citet{g07} and our ionization profiles agree well with those of \citet{y04}.
Both \citet{g07} and \citet{y04} model hydrogen and helium molecular
chemistry in hot Jupiter winds, and \citet{g07} includes contributions
from D, C, N, O, and CH.  \citet{g07} finds that if metals are present
in the wind in solar abundances, they can increase the mass loss rate
by factors of a few by reducing the effectiveness of
H$_3^+$ cooling at depth (below the $\tau = 1$
surface to photoionization).  Again, these details can be considered
buried in our parameter $r_{\rm min}$.

Neither \citet{y04} nor \citet{g07}
considered the effects of Lyman-$\alpha$ line cooling---which we have
demonstrated is important for large UV fluxes---or of line cooling from
metals such as \ion{O}{II}, \ion{O}{III}, and \ion{N}{II}.  If
classic \ion{H}{II} regions are any guide, metal line cooling
could exceed Ly$\alpha$ cooling, lowering hot Jupiter wind temperatures
by up to a factor of 2 \citep{o89}.
Metal abundances in the upper thermosphere are extremely uncertain,
depending on unknown turbulent mixing coefficients (``eddy diffusivities'').
In any case, we do not expect metal line cooling to change our results
qualitatively. It can only lower mass loss rates and
wind velocities somewhat, strengthening our main conclusions.

It is often said that transonic winds are characterized by zero
pressure at infinite distance, while subsonic breezes have finite
pressure at infinity \citep[e.g.,][]{g07}. But in practice, when
considering how the outflow from the planet's dayside interacts with
the outflow from its host star, whether the dayside emits a wind or a
breeze does not require us to examine conditions at infinity. Rather,
we evaluate conditions at the sonic point.  Again, as with our
criterion for hydrodynamic escape, the sonic point serves as
discriminant because once the flow achieves supersonic velocities past
the sonic point, it cannot influence the flow inside the sonic point.
If the total external pressure (ram, thermal, and magnetic) exerted by
the host star's wind is less than the total pressure exerted by the
transonic planetary wind at its sonic point, then the transonic wind
solution obtains.  Otherwise, either the dayside emits a more gentle
breeze or---if the external pressure exceeds some critical value---the
dayside atmosphere does not escape at all but is forced to be in
vertical hydrostatic equilibrium.  To the extent that the host star of
HD 209458b emits a highly variable wind like that of the Sun, we find
that the stellar wind pressure is comparable to the planetary
wind pressure at the sonic point (both are measured in tens of
picobars).  During violent coronal mass ejections, the stellar wind
pressure may overwhelm the planetary wind pressure.  Thus we expect
that dayside winds from hot Jupiters orbiting main-sequence solar type
stars will alternately turn on and off (with a duty cycle that might
well favor the off state).  When the dayside wind is off, we expect
the nightside wind to turn on and pick up the slack---the absorbed UV
energy now being used to power horizontal flows that carry
photoionized plasma to the planet's nightside, which is shielded from
the stellar wind and therefore immune to pressure confinement.

Winds from T Tauri stars have magnetic and ram pressures that are some
six orders of magnitude greater than their main-sequence
counterparts. Were a hot Jupiter orbiting a T Tauri star to also emit
a wind, the pressure at its sonic point would be greater than under
main-sequence conditions, but only by about three orders of magnitude
according to our model.  Thus, a T Tauri stellar wind will entirely quash dayside outflows from orbiting hot Jupiters.  Outflows from the nightside (if the planet is outside the stellar Alfv\'en radius) or along polar field lines (if the planet is inside the Alfv\'en radius) would carry away at most $\sim$$2 \times 10^{12}\gm \s^{-1}
\times 10^7 \yr$ or $\sim$$6\times 10^{-4}$ of the planetary mass.
These quantitative considerations rule out speculations that mass loss
is significant during the host star's youth \citep[e.g.,][]{bsc+04}.

\begin{figure}
\centering
\hbox{\hspace{-0.25in}\scalebox{1.3}{\plotone{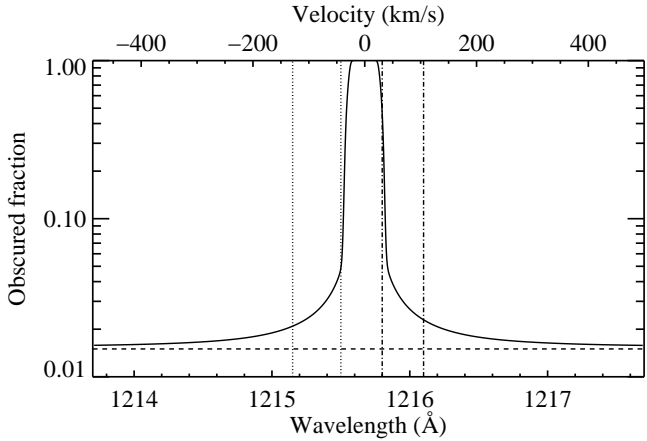}} }
\caption{Maximum fraction of stellar Ly$\alpha$ radiation obscured by
  absorption in the planetary wind, plotted as a function of Doppler
  shifted velocity from line center.  We apply our wind solution for
  the substellar streamline over all $4\pi$ steradians of the planet's
  surface. We assume that the planet is located at mid-eclipse, and
  that its orbital inclination relative to the stellar equator is
  zero.  For each projected distance from the planet's center ranging
  from $R_{\rm p}$ to $8.2 R_{\rm p} = R_{\ast}$, we calculate a
  line-of-sight optical depth, integrating up to a maximum altitude of
  $10 R_{\rm p}$.  We include both natural and thermal line
  broadening.  The ordinate is the fraction of the stellar disk that
  is obscured, accounting only for $e^{-\tau}$ absorption and
  neglecting scattering and wind self-emission.  All of our
  assumptions maximize the obscuration.  Absorption drops sharply
  beyond a few tens of km/s to a constant of 1.5\% (horizontal dashed
  line), the decrement in the visible continuum. The velocity scale of
  $\sim$20 km/s is set by thermal broadening ($\sim$$10$ km/s at
  $10^4\K$), together with the bulk velocity (10--30 km/s at Mach
  numbers of a few). If the wind is emitted only from the dayside
  (nightside), our extinction curve will be valid only at wavelengths
  redward (blueward) of line center.  Absorption by the planetary wind
  does not account for the claimed $\sim$9--15\% decrement in 
  flux from HD 209458b at Doppler equivalent velocities
  near $\pm 100$ km/s, as integrated over the intervals between the
  two dotted lines and between the two dot-dashed lines
  \citep[][]{vld+03,b07}.  We further emphasize this inconsistency in
  Figure \ref{fig-vm2}.}
\label{fig-obscure}
\end{figure}

Returning to the main-sequence case,
to what degree will a hot Jupiter wind absorb stellar Ly$\alpha$
photons and produce an observable transit signature?
Figure \ref{fig-obscure} shows, for standard model
parameters, how the extinction varies with wavelength in the
Ly$\alpha$ line, assuming the planet is in mid-transit and that its
orbit is viewed edge on.  The extinction is integrated across the
entire stellar disk (to perform this integral, we extend our wind
model out to $r = 10 R_{\rm p} \approx 2 R_{\rm Roche}$).
Local line profile functions are
Voigt profiles, which include thermal and natural broadening.  Out to
Doppler shifts of $\pm 30 \km \s^{-1}$, comparable to bulk wind
velocities, the line is essentially black (the Ly$\alpha$ emission
produced by the wind itself is negligible since densities are low enough that the $n=2$ population of neutral hydrogen is well below the LTE value).  At Doppler shifts of $\pm
100 \km \s^{-1}$, the extinction plummets to 2--3\%. This result is at
odds with the observational claim that HD 209458b reduces the stellar
Ly$\alpha$ flux by some 9--15\% at $\pm 100 \km \s^{-1}$ while in transit
\citep{vld+03,b07,vld+08}.

\begin{figure}
\centering
\hbox{\hspace{-0.25in}\scalebox{1.3}{\plotone{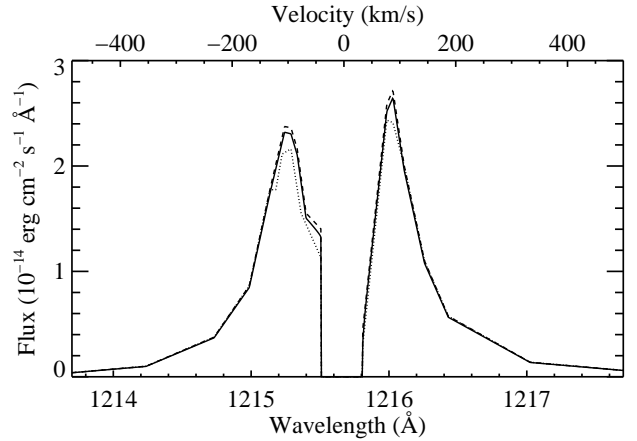}} }
\caption{Observed out-of-transit (dashed line) and in-transit (dotted
line) spectra, reproduced from Figure 2 of \citet{vld+03}. In the line ``core''
from -42 to +32 km/s, where interstellar absorption is strong, we 
set the flux to zero. Our theoretical in-transit
spectrum (solid line) is computed by attenuating
the observed out-of-transit spectrum according to our extinction
curve in Figure \ref{fig-obscure}. Theory predicts
substantially less absorption than is claimed to be observed.
However, the spectrally
unresolved measurements of \citet[][]{vdl+04}, taken at a different
epoch and in principle pertaining to much smaller Doppler
equivalent velocities than were implicated by \citet{vld+03},
may well be due to
the wind. Integrated over the line profile outside the core,
our model predicts a flux decrement of 2.4\%, close to that observed
by \citet{vdl+04}.}
\label{fig-vm2}
\end{figure}

To drive this point home, we present in Figure \ref{fig-vm2} the
predictions of our model for the Ly$\alpha$ line during transit,
computed by combining the out-of-transit line spectrum from
\citet[][taken from their Figure 2]{vld+03} with our extinction curve
(Figure \ref{fig-obscure}). Over the wavelength intervals used by
\citet{vld+03}, our model produces a flux decrement of 2.9\%.  Shown
for comparison is the observed in-transit line spectrum from
\citet{vld+03}. Agreement between model and observation is poor: the
model spectrum is hardly absorbed at large velocities, by contrast to
the observed spectrum.

What could explain this discrepancy between our model and the observations?  Three possibilities present themselves: (1) Our model underestimates the column of neutral hydrogen generated by the wind during transit by a factor of $\sim$3--5; (2) physics that is missing from our model generates a population of neutral hydrogen moving at velocities larger than the bulk velocity of the wind; or (3) the observed flux decrement at $\pm$100 km/s is due to some combination of geocoronal emission and intrinsic stellar variability (see also Ben-Jaffel 2007): processes that, if observed over long enough times, should not be correlated with the orbital phase of the planet.  We now comment on possibilities (1) and (2).  

Our wind model generates a Ly$\alpha$ flux decrement of $\sim$3\% at Doppler-equivalent velocities of $\pm 100$km/s due to naturally broadened, Lorentzian line wings.  If the column of neutral hydrogen traversed by stellar Ly$\alpha$ photons along lines of sight located 5--10$R_{\rm p}$ from the planet were larger by a factor of 3--5, then absorption generated by the line wings might generate the 9--15\% absorption observed at $\pm 100$km/s.  Ben-Jaffel (2008) argues that this is the case using the wind profiles of Garcia-Munoz (2007), who calculates that the wind is $\sim$30\% neutral at $\sim$5$R_{\rm p}$.  At these distances, we find a neutral fraction of $\sim$13\% for our base case and $\sim$6\% for parameters matching those used in Garcia-Munoz (2007), though in other respects our wind solutions largely agree.  Our ionization profiles are in better agreement with the solutions of Yelle (2004), which do not generate sufficient absorption in the Lorentzian line wings to match observations (Ben-Jaffel 2008).  The reason for these differences merits further attention. As previously noted, 1D wind solutions such as those in this paper, Yelle (2004), and Garcia Munoz (2007) break down at planetocentric distances greater than $\sim$5--10$R_{\rm p}$.  Multidimensional calculations could yield a larger neutral column (J.~Stone, personal communication).

Alternatively, is there some qualitative physics that our model is missing that could generate a population of neutral H atoms moving at velocities of $\sim$100 km/s?  
An appeal might be made to interaction of the
planetary wind with the stellar wind: in the bow shock, neutral
hydrogen from the planet mixes with fast moving stellar plasma and
might be accelerated to large blueshifted velocities (see the
simulations by \citealt{sbp92} of colliding stellar winds).  But no
similar argument can be made for the observed redshifted
absorption---which, according to \citet{b07}, sometimes appears
stronger than the blueshifted absorption (see transit B2 in his Figure
3b).  Appeals to radiation pressure \citep{vld+03} founder for the
same reason. \citet{g07} suggests turbulence in the planetary wind
itself as a way to broaden the line. But to generate the large
velocities observed, the energy in such turbulence would need to
exceed the thermal and bulk kinetic energies in the mean flow by a
factor of $\sim$100.  Such energy requirements seem
insurmountable. Coriolis forces can turn streamlines that are
initially in the plane of the sky into our line of sight, producing
both redshifted and blueshifted gas. But the time required for gas to
reach $\sim$10$R_{\rm p}$, the size scale probed by the transit
measurements, is too short to produce line-of-sight velocities of
$\sim$100 km/s.  In short, because the planetary wind stalls at the
bow shock as it blows towards the star, stellar gravity cannot
accelerate the flow to large redshifted velocities.

More promising is the possibility that charge-exchange between stellar wind protons and H atoms in the planetary wind could generate high velocity neutrals.  Holmstrom et al. (2008) find that H atoms will be accelerated by charge-exchange to velocities of $\sim$100 km/s on the assumption that the stellar wind interacts directly with the planetary magnetosphere at $\sim$4$R_{\rm p}$ and that neutrals from the planet have been lifted to that height. In neglecting the planetary magnetic field, we have modeled planetary winds whose pressures exceed those of their magnetic fields.  In this case, charge-exchange in the shock between the stellar and planetary winds might likewise accelerate H atoms \citep[see, e.g.,][for a discussion of charge-exchange in the bow shock of an infalling comet]{rfs+98}.  Hot Jupiter magnetic field strengths are uncertain but magnetospheres may compete with planetary winds for the dominant source of pressure at high altitude (\S\ref{sec-breeze}).  Whether the stellar wind forms a shock with the planetary wind or with the planet's magnetosphere may vary from system to system.

We conclude that although UV radiation from main-sequence stars can drive hot Jupiter winds with mass loss rates of $\sim$$10^{10}$ g/s, the source of the observed absorption detected at Doppler-equivalent velocities of $\pm$100 km/s in HD 209458b remains uncertain, with several possible candidates.  What does our model predict for spectrally unresolved measurements?  \citet{vdl+04} collect light over all Doppler equivalent velocities and indicate a wavelength-integrated flux decrement of
$5\pm 2$\%.  We take the out-of-transit
line spectrum from Figure 2 of \citet{vld+03} and reduce it according
to the obscured fraction computed in our Figure
\ref{fig-obscure}. Integrating over the range 1213.7--$1217.7 \AA$, and
excluding the line core between 1215.5 and $1215.8\AA$ (velocities
between -42 km/s and +32 km/s) inside of which interstellar absorption
practically extinguishes the line \citep[see Figure 1 of][]{vld+03},
we compute a wavelength-integrated Ly$\alpha$ flux decrement of 2.4\%
(compare to the flux decrement in the visible continuum, 1.5\%).  As a
check, we apply our procedure to the observed in-transit spectrum of
\citet{vld+03}, finding a flux decrement of 5.3\%, in good agreement
with the 5.7\% quoted by \citet{vdl+04}. While our model flux decrement
of 2.4\% is sensitive to our assumed outer cut-off radius ($10 R_{\rm
  p}$), and is uncertain because our model breaks down there (we
neglect the full stellar gravity field and Coriolis forces), it is
nevertheless close enough to the observed decrement of $5\pm 2$\%
\cite[][]{vdl+04} that the spectrally unresolved measurements may well
be probing a planetary outflow. We look to the Hubble Space Telescpe to reproduce this signature of a hot Jupiter wind after STIS is repaired or the Cosmic Origins Spectrograph (COS) is installed.

\acknowledgements This work was supported by a Berkeley Atmospheric
Sciences Center Fellowship and an American Association of University
Women Fellowship, both held by R.M.-C., and by Hubble Space Telescope
Theory Grant HST-AR-11240.01-A.  We thank Jon Arons for a careful
reading of the manuscript and for helping us appreciate some aspects
of magnetospheric physics.  We also thank Lofti Ben-Jaffel, Steven Cranmer, Doug Lin, Geoff Marcy,
Eliot Quataert, John Raymond, Jim Stone, and Josh Winn for helpful and
encouraging conversations. Finally we are grateful to the referee
for an authoritative and thorough report that led to substantive
improvements in our work, especially regarding the interpretation of the
observations.

\begin{appendix}
\section{Sensitivity to Boundary Conditions}\label{sec-bcsense}

We demonstrate that our solution is insensitive to
our choices for BC3 through BC6 (\S\ref{sec-bcs}).
That is, we show that our standard model represents
a ``quasi-unique'' solution
that hardly changes over large and physically realistic
regions of input parameter space.
We also show that while our solution does depend sensitively
on $r_{\rm min}$ (which does not enter as a formal boundary
condition but represents instead a global scale factor),
that parameter is known sufficiently well
that it introduces no more than a factor of 2 uncertainty in our
determination of the mass loss rate.

\begin{figure}
\begin{minipage}[b]{0.45\linewidth}
\centering
\scalebox{1.2}{\plotone{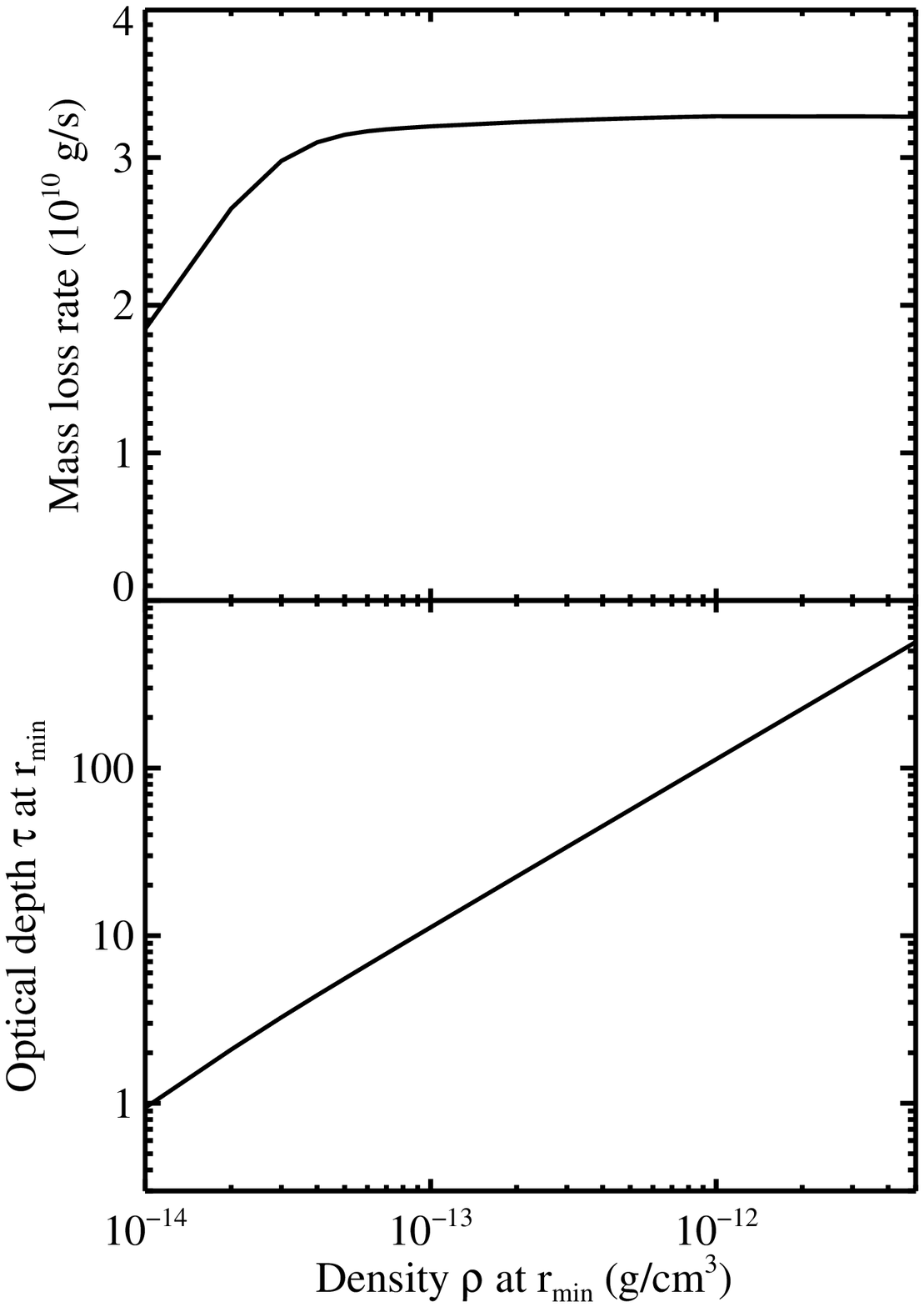}}
\caption{Mass loss rate $\dot{M}$ and $\tau(r_{\rm min})$ as a
  function of boundary condition BC3: $\rho(r_{\rm min})$.
  The other boundary conditions BC4--BC6 are kept fixed at
  their standard values, and $r_{\rm min} = R_{\rm p}$.
  To calculate $\dot{M}$, we apply our 1D solution over a full $4\pi$
  steradians.  As long as we choose $\rho(r_{\rm min})$ sufficiently large
  that $\tau(r_{\rm min}) \gg 1$, $\dot{M}$ is
  insensitive to our choice.}
\label{fig-bcrho}
\end{minipage}
\hspace{0.5cm}
\begin{minipage}[b]{0.45\linewidth}
\centering
\scalebox{1.2}{\plotone{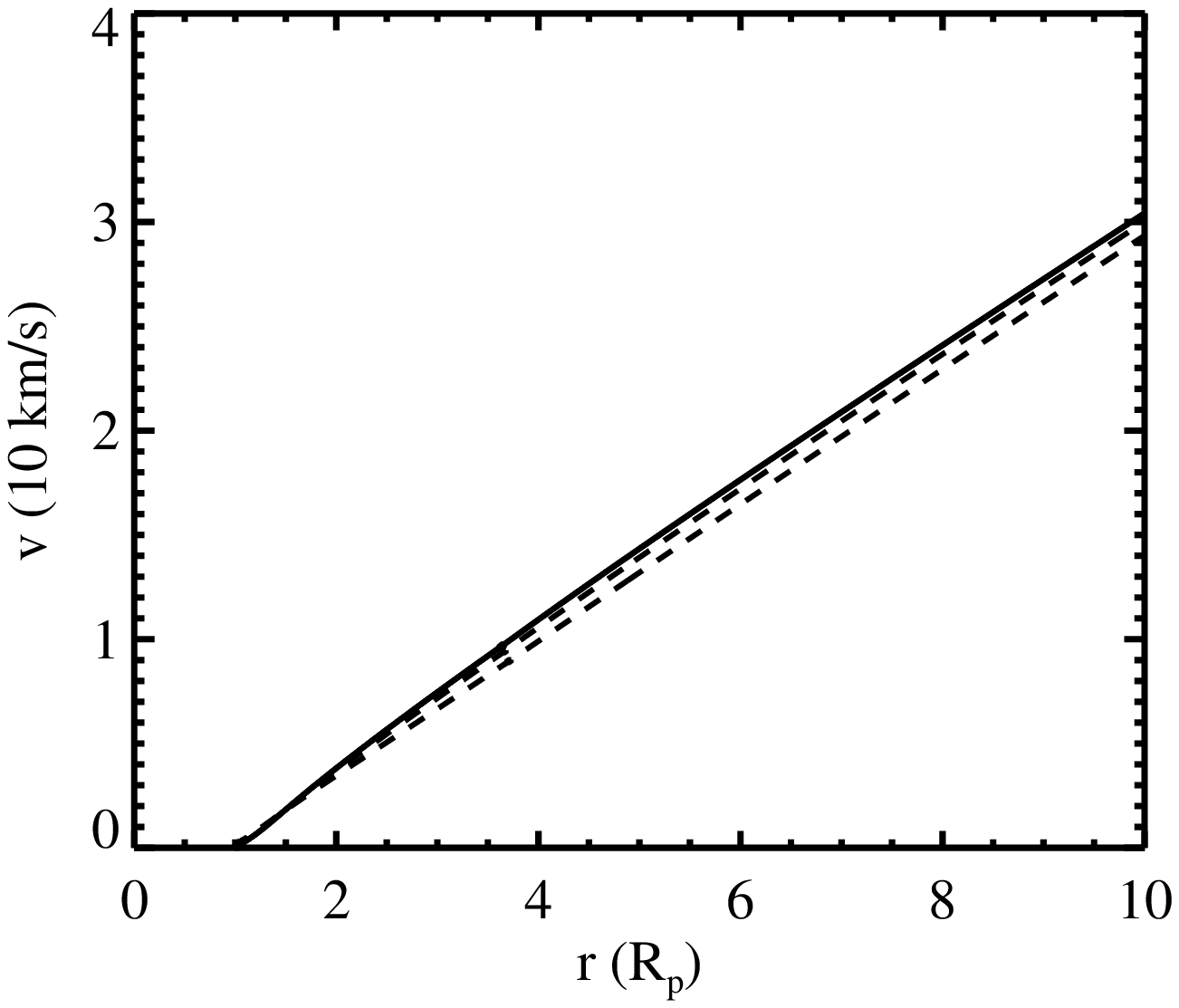}}
\caption{Wind velocity as a function of radius for different choices
  of BC3: $\rho(r_{\rm min})$.  For $\rho(r_{\rm min}) = 4\times
  10^{-13}$ to $5\times 10^{-12}$ g/cm$^3$, the solutions are
  indistinguishable from the solid line.  For lower densities
  $\rho(r_{\rm min}) = 2\times 10^{-14}$ (top dashed line) and
  $1\times 10^{-14}$ (bottom dashed line), $\tau(r_{\rm min})$ is not
  $\gg 1$ and the profiles are sensitive to BC3.}
\label{fig-bcrho2}
\end{minipage}
\end{figure}

Regarding BC3,
Figures \ref{fig-bcrho} and \ref{fig-bcrho2} show that as long as the
base density $\rho(r_{\rm min})$ is large enough that
$\tau(r_{\rm min}) \gg 1$, the solution is insensitive to $\rho(r_{\rm
  min})$. Our standard value $\rho(r_{\rm min}) = 4 \times 10^{-13}
\gm \cm^{-3}$ gives $\tau(r_{\rm min}) = 50$ and so satisfies this requirement.

\begin{figure}
\begin{minipage}[t]{0.45\linewidth}
\centering
\scalebox{1.2}{\plotone{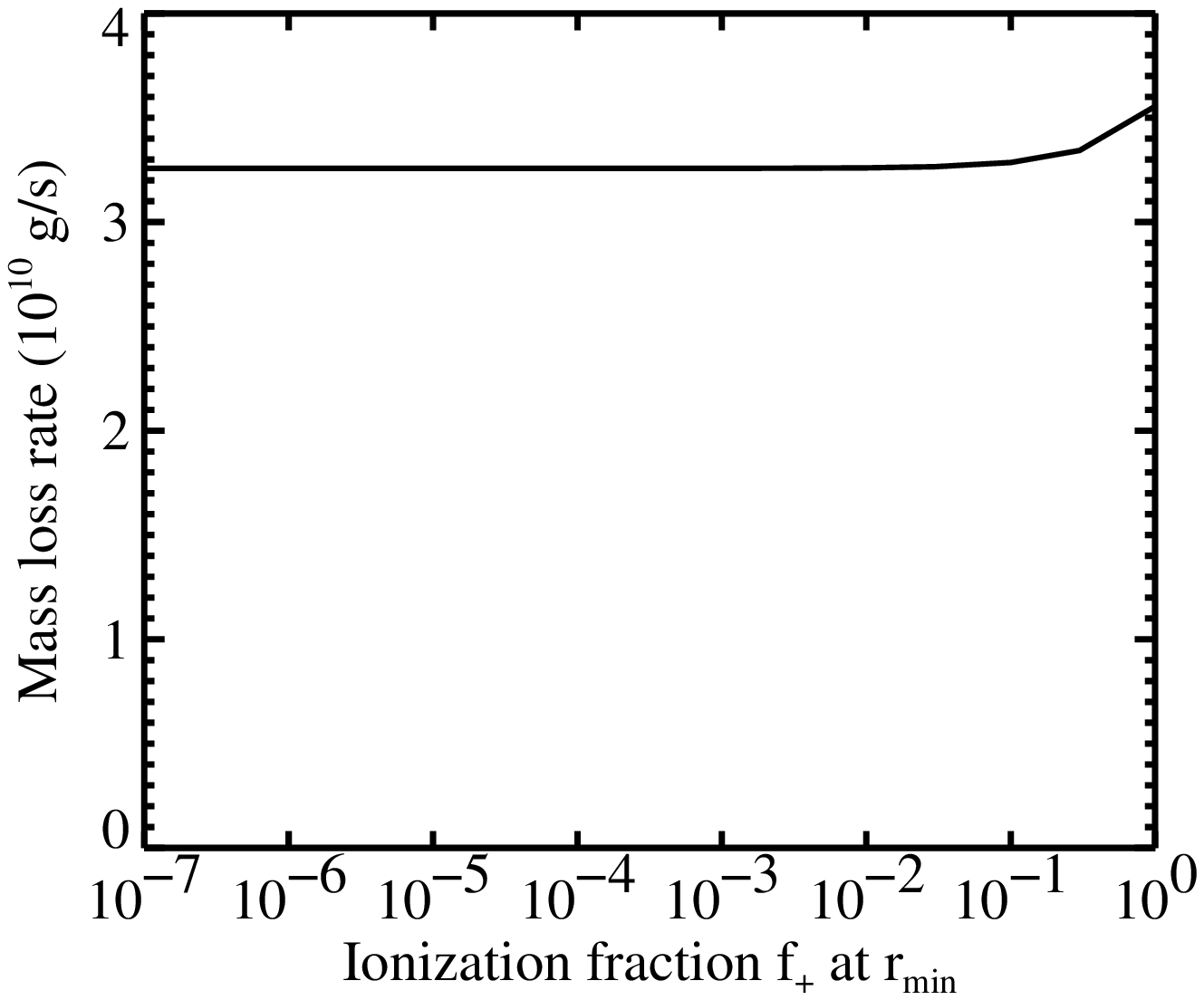}} 
\caption{Mass loss rate $\dot{M}$ as a function of BC4: $f_+(r_{\rm min})$.
  The other boundary conditions BC3, BC5, and BC6 are kept fixed at their
  standard values,
  and $r_{\rm min} = R_{\rm p}$.  To calculate
  $\dot{M}$, we apply our 1D solution over $4\pi$ steradians.
  As long as $f_+(r_{\rm min}) \ll 1$, $\dot{M}$ is insensitive to
  $f_+(r_{\rm min})$.}
\label{fig-bcfp}
\end{minipage}
\hspace{0.5cm}
\begin{minipage}[t]{0.45\linewidth}
\centering
\scalebox{1.2}{\plotone{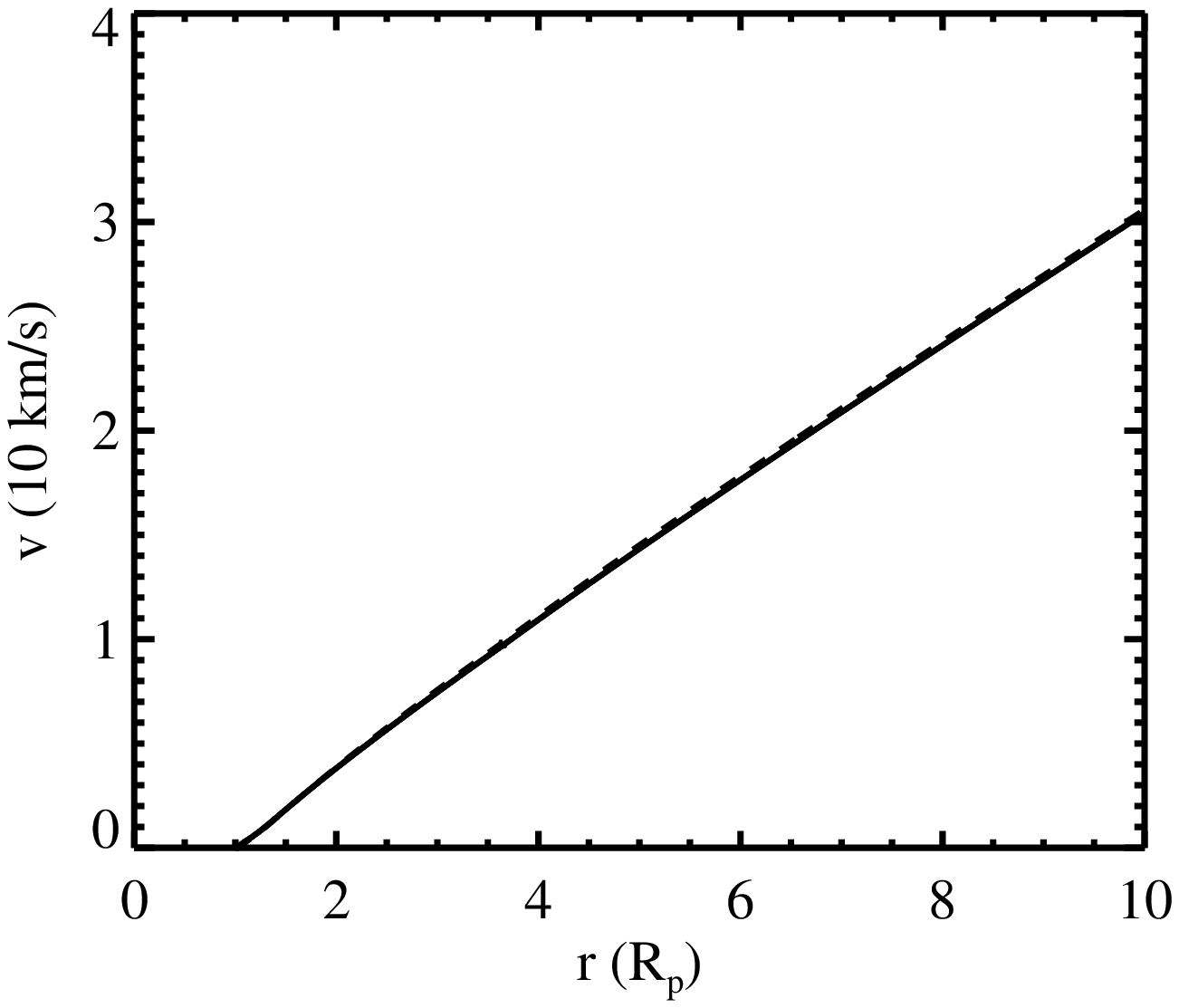}} 
\caption{Wind velocity as a function of radius for different choices
  of BC4: $f_+(r_{\rm min})$.  For $f_+(r_{\rm min}) = 10^{-7}$ to
  $10^{-1}$, the solutions are indistinguishable (solid line). For
  $f_+(r_{\rm min}) = 1$ (dashed line), the solution differs only slightly.
  For that solution, $f_+$ drops to a small value at $r$ just above
  $r_{\rm min}$.}
\label{fig-bcfp2}
\end{minipage}
\end{figure}

For BC4, we set $f_+(r_{\rm min})$ to an arbitrary number $\ll 1$.
When $\tau(r_{\rm min}) \gg 1$,
$f_+(r_{\rm min}) \ll 1$ and our solution is not
sensitive to its exact value (Figures
\ref{fig-bcfp} and \ref{fig-bcfp2}).  In addition, we have verified that
our solution does not change if
we replace BC4 with the requirement that photoionizations balance radiative
recombinations at $r_{\rm min}$.

\begin{figure}
\begin{minipage}[t]{0.45\linewidth}
\centering
\scalebox{1.2}{\plotone{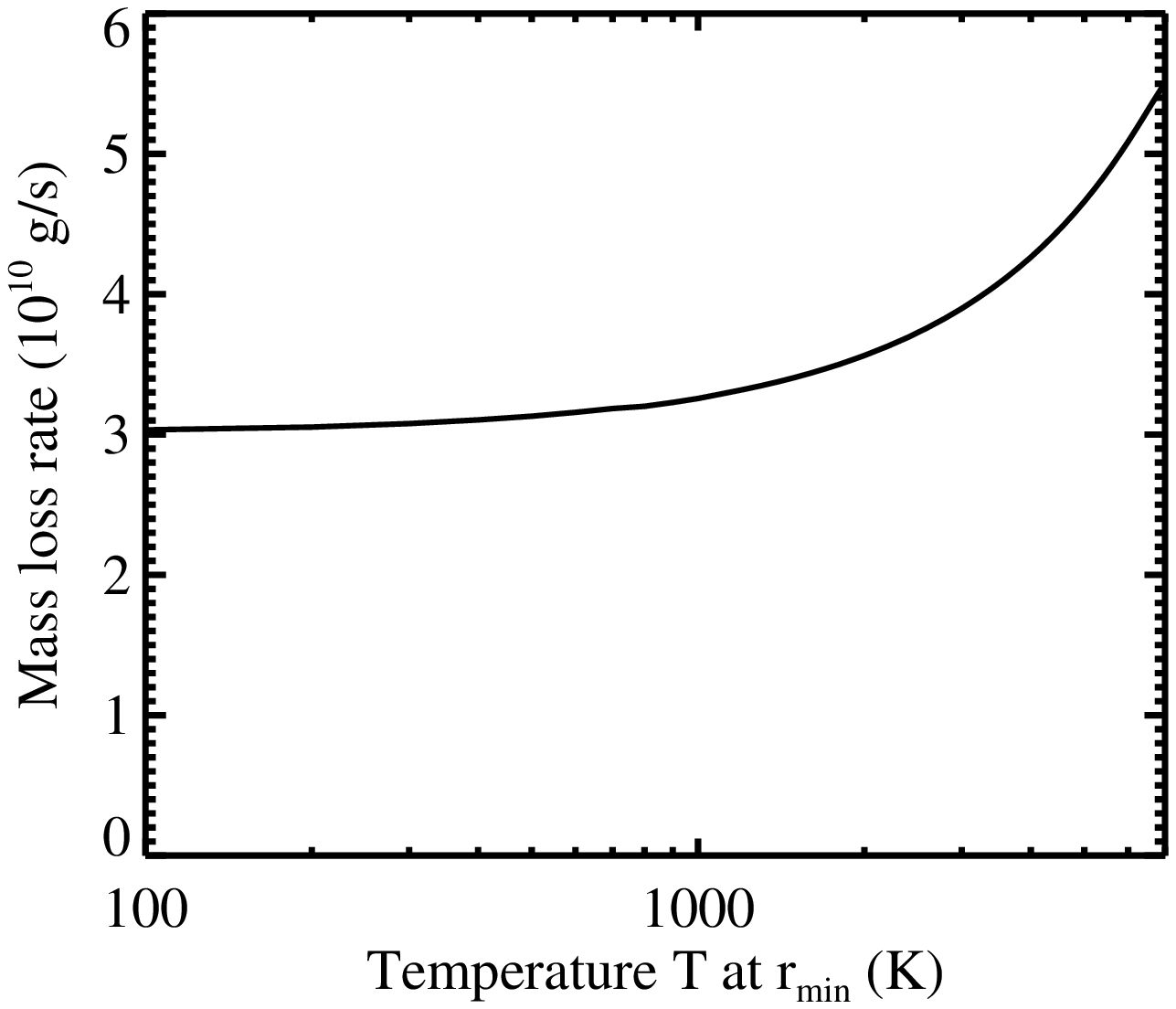}} 
\caption{Mass loss rate $\dot{M}$ as a function of BC5: $T(r_{\rm
    min})$. Boundary conditions BC4 and BC6 are kept fixed at their standard
  values, 
  and $r_{\rm min} = R_{\rm p}$.  For BC3, we take $\rho(r_{\rm min}) =
  4\times 10^{-13}$ g/cm$^3$ for $T(r_{\rm min}) \ge 800$ K, $8\times 10^{-13}$
  g/cm$^3$ for 800 K $>T(r_{\rm min}) > 100$ K, and $3\times 10^{-12}$ g/cm$^3$ for
  $T(r_{\rm min}) = 100$ K.  These adjustments in $\rho(r_{\rm min})$ are made
  to maintain $\tau(r_{\rm min}) \gg 1$ (see Figures \ref{fig-bcrho}
  and \ref{fig-bcrho2}).   To calculate $\dot{M}$, we apply our 1D
  solution over $4\pi$ steradians.  As long as $T(r_{\rm min})
  \ll 10^4$ K, $\dot{M}$ is insensitive to $T(r_{\rm min})$.}
\label{fig-bcT}
\end{minipage}
\hspace{0.5cm}
\begin{minipage}[t]{0.45\linewidth}
\centering
\scalebox{1.2}{\plotone{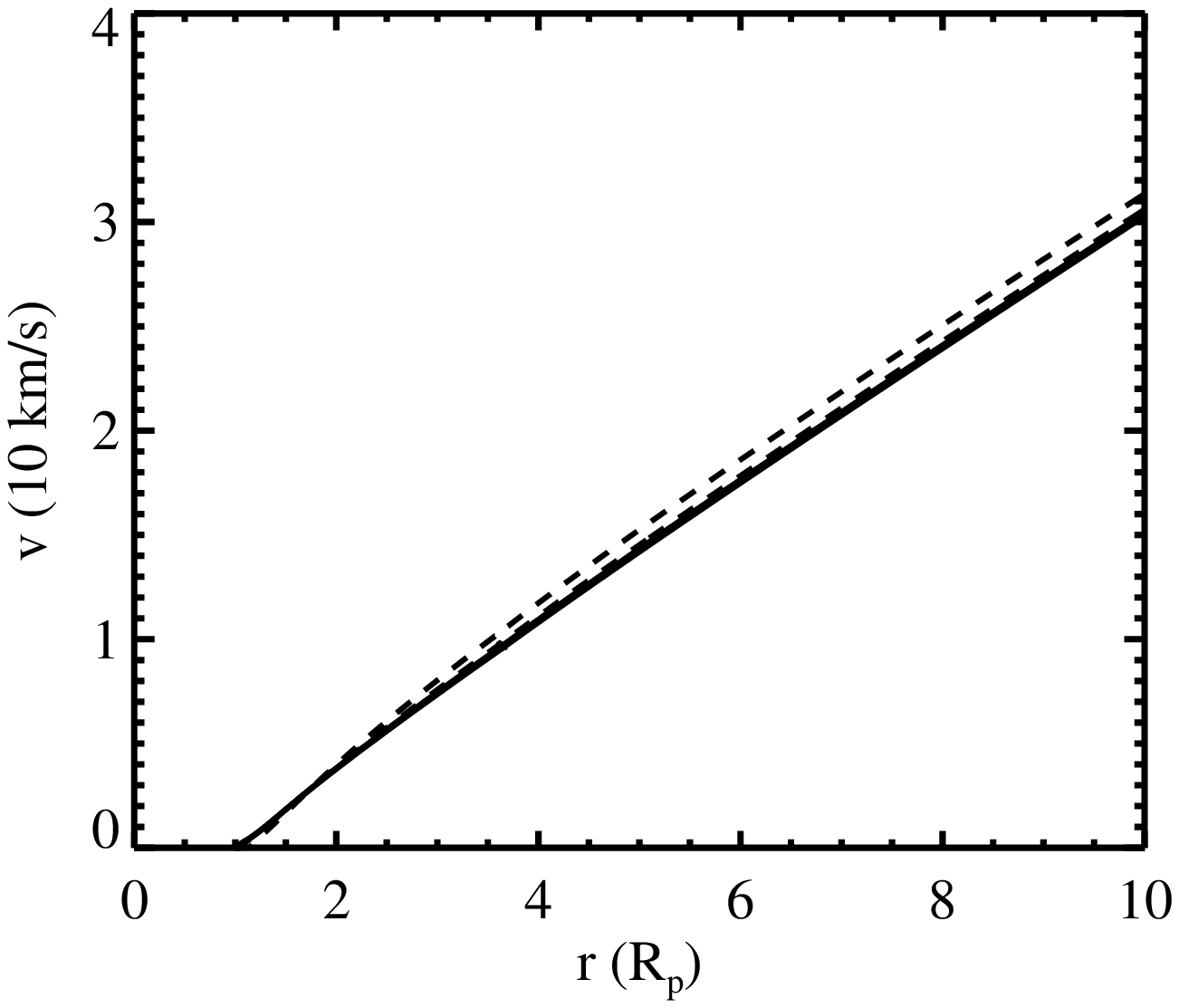}} 
\caption{Wind velocity as a function of radius for different choices
  of BC5: $T(r_{\rm min})$.  For $T(r_{\rm min}) = 100$ to 1000 K, the
  solutions are practically indistinguishable from one another.
  For higher temperatures $T(r_{\rm
    min}) = 2000$ K (lower dashed line) and 6000 K (upper dashed line),
  the wind profiles change.}
\label{fig-bcT2}
\end{minipage}
\end{figure}

For BC5, we have chosen $T(r_{\rm min}) = 1000$ K for our standard value.
As demonstrated in Figures \ref{fig-bcT} and
\ref{fig-bcT2}, our solution is insensitive to this choice as long as
$T(r_{\rm min}) \ll 10^4$ K (the temperature such that thermal velocities
are comparable to the local escape velocity).
All models 
of hot Jupiter atmospheres at depth \citep[e.g.,][]{bsh03} have
this property.

\begin{figure}
\begin{minipage}[t]{0.45\linewidth}
\centering
\vspace{-0.1in} 
\scalebox{1.2}{\plotone{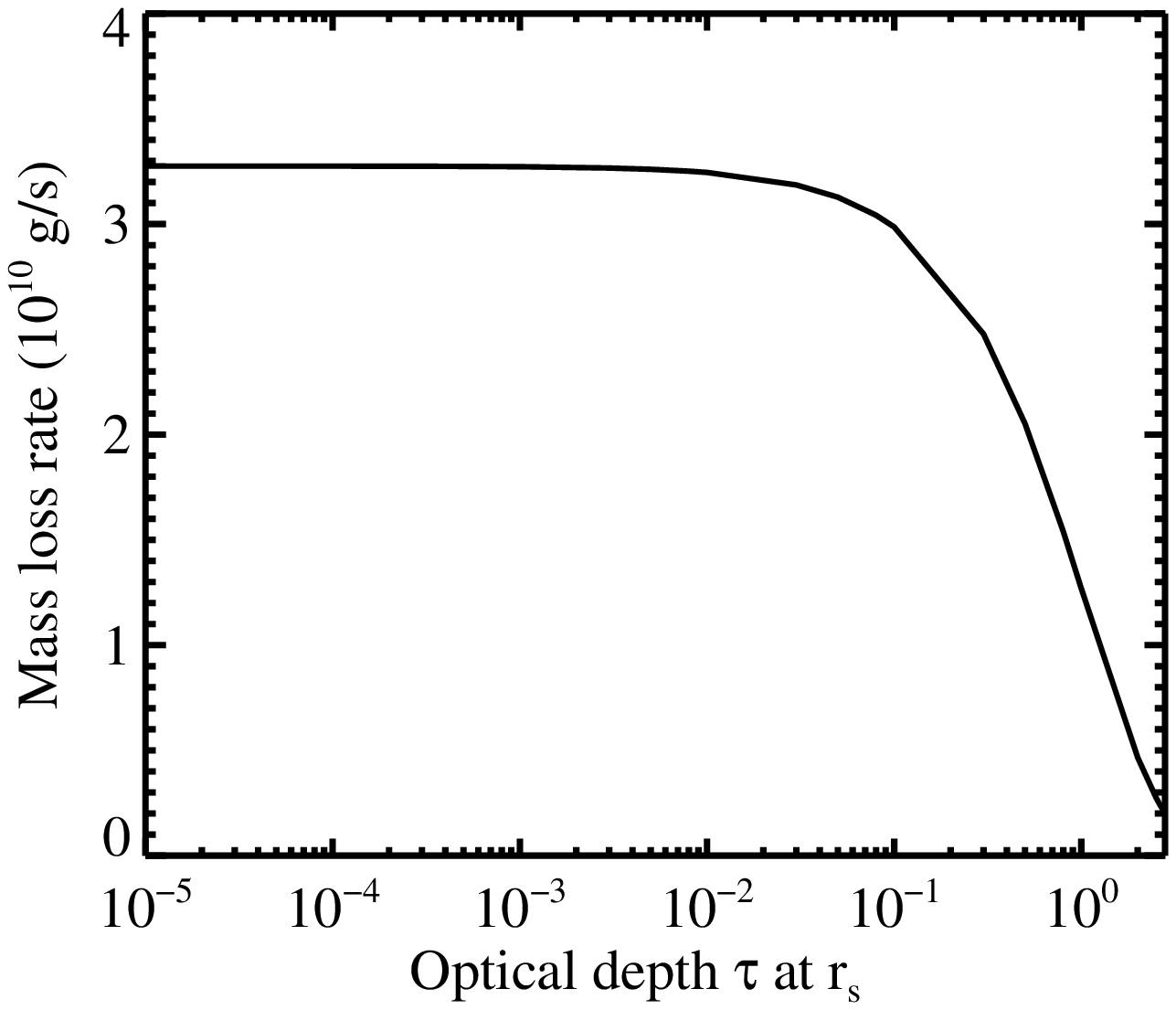}} 
\caption{Mass loss rate $\dot{M}$ as a function of BC6: $\tau(r_{\rm
    s})$.  The other boundary conditions BC3--BC5 are kept fixed at their
  standard values,
  and $r_{\rm min} = R_{\rm p}$.
  To calculate $\dot{M}$, we apply our 1D solution over $4\pi$
  steradians.  Provided $\tau(r_{\rm s}) \ll 1$---which it self-consistently
  is for
  our quasi-unique solution---$\dot{M}$ is
  insensitive to $\tau(r_{\rm s})$.  
}
\label{fig-bctau}
\end{minipage}
\hspace{0.5cm}
\begin{minipage}[t]{0.45\linewidth}
\centering
\vspace{-0.1in} 
\scalebox{1.2}{\plotone{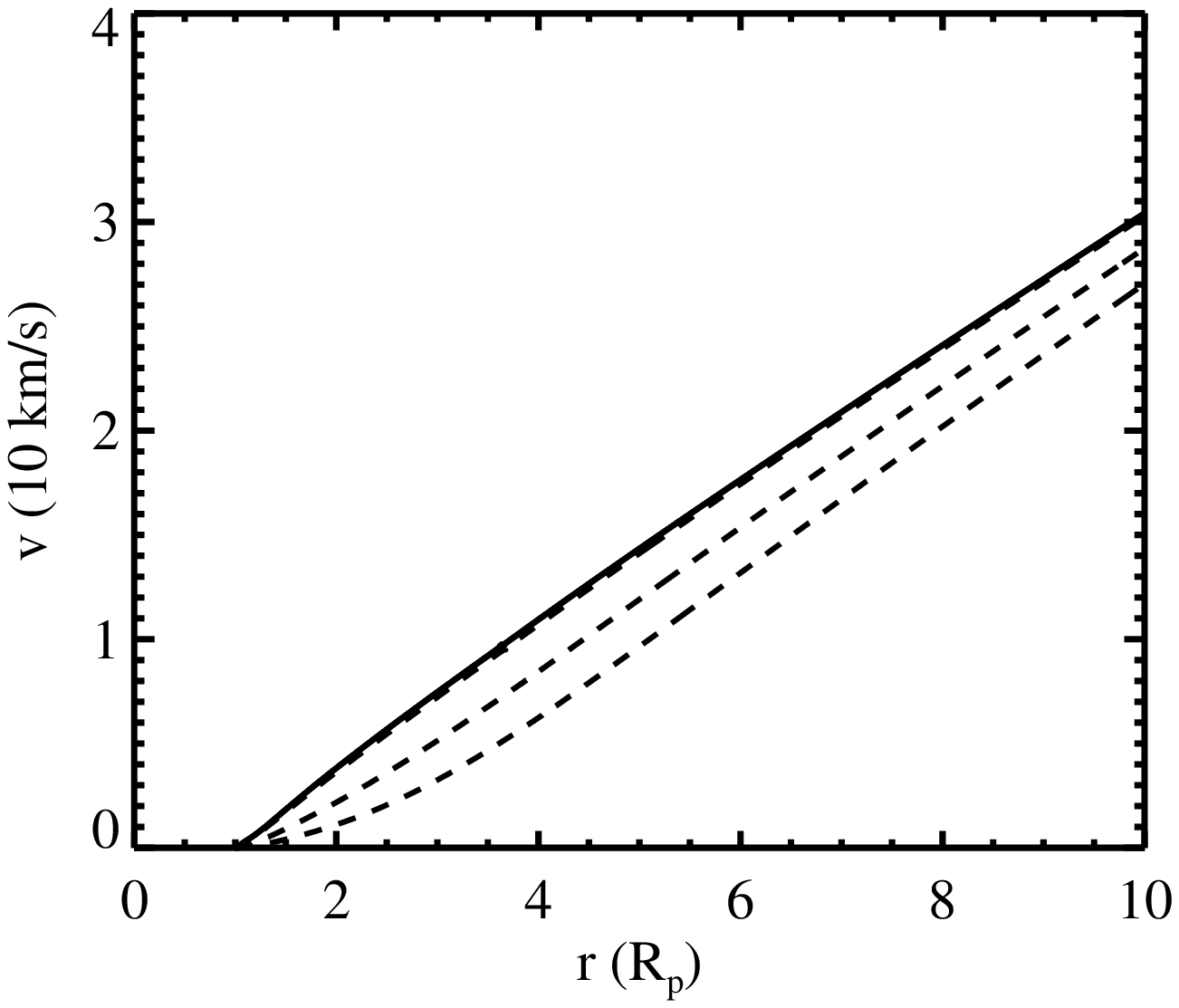}} 
\caption{Wind velocity as a function of radius for different choices
  of BC6: $\tau(r_{\rm s})$.  For $\tau(r_{\rm s}) = 10^{-5}$ to $10^{-2}$,
  the solutions are indistinguishable from the solid line.  For larger
  $\tau(r_{\rm s}) = 0.1$ (top dashed line), 1.0 (middle dashed line),
  and 2.0 (bottom dashed line), the solution changes. But $\tau(r_{\rm s}) < 1$
  is demanded by the quasi-unique solution that appears when varying
  the other boundary conditions BC3--BC5.}
\label{fig-bctau2}
\end{minipage}
\end{figure}

Regarding BC6, Figures
\ref{fig-bctau} and \ref{fig-bctau2} demonstrate that
our solution is independent of
our choice for $\tau(r_{\rm s})$ as long as it is $\ll 1$.
This requirement is satisfied by the quasi-unique solution
that appears when varying the other boundary conditions BC3--BC5.
Thus, we have shown that our quasi-unique solution self-consistently demands
$\tau(r_{\rm s})\ll 1$.
The condition that $\tau(r_{\rm s}) \ll 1$ is also reasonable because
the optical depth of material outside the planet's Roche lobe
should be small.

\begin{figure}
\centering
\vspace{0.0in}  
\scalebox{0.51}{\plotone{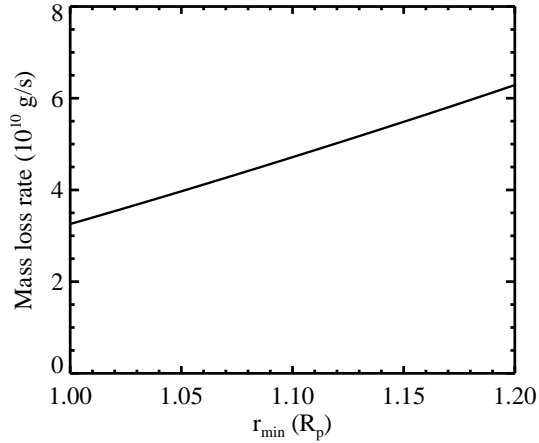}} 
\caption{Mass loss rate $\dot{M}$ as a function of model parameter
  $r_{\rm min}$.  
  Boundary conditions BC3--BC6 are held fixed at their standard values.
  To calculate $\dot{M}$, we apply our 1D solution over a full $4\pi$
  steradians. Though the dependence of $\dot{M}$ on $r_{\rm min}$ is strong,
  the latter is known to within 10\% (see text).}
\label{fig-bcrmin}
\end{figure}

Finally, because we do not solve for the structure of the planet's atmosphere
below $r_{\rm min}$, we are not sure whether our standard value for
$r_{\rm min} = R_{\rm p} \equiv 10^{10} \cm$ corresponds correctly
to our adopted base conditions.
In other words, we cannot say
with certainty whether our adopted base temperature of 1000 K
and base density of $4 \times 10^{-13} \gm \cm^{-3}$ are indeed
reached at our assumed radius of $r_{\rm min} = 10^{10} \cm$. 
The value of the base radius is important since it helps determine
the effective planetary cross section for absorption of stellar radiation
and the local planetary gravity, both of which affect
the mass loss rate---see Equation (\ref{eqn-energylimited}).
Figure \ref{fig-bcrmin} shows
how our results depend on 
$r_{\rm min}$. The mass loss rate $\dot{M}$ is indeed very sensitive
to $r_{\rm min}$, changing by as much as a factor
of 2 when $r_{\rm min}$ changes by just 20\%.
This is consistent with the $\dot{M}_{\rm e-lim} \propto R_{\rm p}^3$
scaling found in Equation (\ref{eqn-energylimited}).
However, our uncertainty in $r_{\rm min}$ is only about 10\%.
We should choose a value for $r_{\rm min}$ that lies between the
1-bar radius (say) and the radius where the bulk
of the stellar UV photons are absorbed. From the
estimates made at the beginning of \S\ref{sec-model},
$1 \lesssim r_{\rm min}/R_{\rm p} \lesssim 1.1$.
Therefore in practice the uncertainty in $\dot{M}$
due to our uncertainty in $r_{\rm min}$ amounts to no more than a factor of 2.

\section{Other Cooling and Ionization Mechanisms}\label{sec-other}

The extra cooling mechanisms we considered and found
to be negligible include collisional ionization 
\begin{equation}\label{eqn-lambdacol}
\Lambda_{\rm col} = -1.3\times 10^{-21}n_+n_0T^{1/2} e^{-157809\K/T} \erg \cm^{-3} \s^{-1}
\end{equation}
\citep{b81},
recombination radiation (appropriate for Case A and thus an overestimate)
\begin{equation}
\Lambda_{\rm rec} = -2.85 \times 10^{-27}T^{1/2}(5.914-0.5\ln{T}+0.01184 T^{1/3}) n_+^2 \erg \cm^{-3} \s^{-1}
\end{equation}
\citep{b81}, 
free-free emission
\begin{equation}\label{eqn-lambdafree}
\Lambda_{\rm ff} = -1.426\times10^{-27}g_{\rm ff}T^{1/2}n_+^2 \erg \cm^{-3} \s^{-1}
\end{equation}
where $g_{\rm ff} \approx 1.3$ is the Gaunt factor \citep{s78},
and conduction
\begin{equation}
\Lambda_{\rm cond} = \frac{1}{r^2}\frac{\partial}{\partial r}\left(r^2 \kappa \frac{\partial T}{\partial r}\right) 
\end{equation}
where the thermal conductivity $\kappa = 4.45\times 10^4 (T/10^3\K)^{0.7}$ erg cm$^{-1}$ K$^{-1}$ s$^{-1}$ \citep{wdw81}. In equations (\ref{eqn-lambdacol})--(\ref{eqn-lambdafree}), densities are in $\cm^{-3}$ and temperatures are in K.
Note that conduction can either cool or heat gas locally (and indeed both
signs are observed---see Figures \ref{fig-450terms} and \ref{fig-5e5terms}).

We also considered how collisional ionizations change the ionization balance.
The collisional ionization rate is given by dividing (\ref{eqn-lambdacol})
by --13.6 eV \citep{b81}. We found this contribution to be negligible.

\section{Escape of Ly$\alpha$ Cooling Radiation}\label{sec-escape}
To act as an effective coolant, Ly$\alpha$ photons must be able to
escape the wind. But the wind is optically thick
to Ly$\alpha$ photons. Radiative cooling is thwarted if
before the photons escape by resonant scattering,
they excite H atoms that subsequently
undergo collisional de-excitation, converting
photon energy back into heat. We show here that this is
not a significant effect.
 
Line photons escape by frequency redistribution: scattering into line wings
where the Ly$\alpha$ optical depth $\tau_{{\rm Ly}\alpha}$ is much reduced.
The number of scatterings
$N_{\rm scat}$ required for a photon to escape is given approximately
by the inverse of the probability $P_{\rm scat}$
that an excited atom emits the photon at a frequency such that
$\tau_{{\rm Ly}\alpha} < 1$. We estimate this probability as
$$
P_{\rm scat} \sim 2\int_{\nu_1}^{\infty} \phi(\nu) d\nu
$$
where $\phi$ is the Voigt line profile function, accounting for
natural and thermal broadening at $T = 10^4\K$, and
$\nu_1$ is the frequency for which
$\tau_{{\rm Ly}\alpha} = 1$,
blueward of line center. The frequency $\nu_1$ is such that
$$
\phi(\nu_1) = \frac{\phi({\rm line \, center})}{\tau_{{\rm Ly}\alpha}({\rm line \,center})}
$$
where
$$
\tau_{{\rm Ly}\alpha}(\rm line\, center) \sim \frac{\sigma_{{\rm Ly}\alpha}({\rm line\,center})}{\sigma_{\nu_0}} \, \tau \sim 3 \times 10^4 \, \tau
$$
and $\sigma_{{\rm Ly}\alpha}({\rm line\,center}) = 6 \times 10^{-14} \cm^2$
is the absorption cross section at line center.
For a photoionization optical depth $\tau \sim 1$, we find
that $N_{\rm scat} \sim P_{\rm scat}^{-1} \sim 1 \times 10^4$.
This estimate neglects differential bulk velocities in the wind,
which tend to decrease $N_{\rm scat}$,
and the random 3D directions with which photons are scattered,
which tends to increase $N_{\rm scat}$. Both effects are expected to be
of order unity.

While the photon is
being scattered, it spends a time $t_{\rm ex} \sim N_{\rm scat}A_{21}^{-1}$
``locked'' inside an excited H atom (in the form of electron excitation
energy), where $A_{21} = 6.3 \times 10^8 \s^{-1}$ is the
Einstein A coefficient. By comparison, if we assume
that collisional de-excitations are dominated by fast thermal electrons,
the time required for an excited H atom
to experience a collisional de-excitation
is $t_{\rm col} \sim (n_+ \sigma_{\rm en}
v_{\rm th,e})^{-1}$, where $\sigma_{\rm en} \sim 10^{-15} \cm^{2}$
is the electron-neutral de-excitation cross section
and $v_{\rm th,e} \sim 400 \km \s^{-1}$ is the electron thermal
velocity.
In Figure \ref{fig-450nfptau}, we
show that $n_+ \lesssim 10^8 \cm^{-3}$ for $\tau \lesssim 1$
for standard model parameters. Putting it all together,
we find that $t_{\rm ex}/t_{\rm col} \lesssim 6 \times 10^{-5}$.
Thus the photon readily diffuses out of the wind before
becoming thermalized. Moreover, the photon spends such a small
fraction of its time locked inside an excited H atom that outward
advection of gas does not change this conclusion.

\end{appendix}

\bibliographystyle{apj}

\end{document}